\PassOptionsToPackage{pdfpagelabels=false}{hyperref}
\documentclass[useAMS,usenatbib]{mnras}
\pdfoutput=1

\usepackage[pdftex]{graphicx}
\usepackage{amsmath}
\usepackage{natbib}
\usepackage[usenames,dvipsnames,svgnames,table]{xcolor}
\usepackage{times}
\usepackage{amssymb}	
\usepackage{aasmacros}
\usepackage{units}

\newcommand{\Msun}{\mathrm{M_{\sun}}}

\newcommand{\kms}{$\rmn{km\,s^{-1}}$} 
\newcommand{\vmax}{$V_{\rmn{max}}$~}
\newcommand{\rmax}{$r_{\rmn{max}}$~}

\def\gsim{ \lower .75ex \hbox{$\sim$} \llap{\raise .27ex \hbox{$>$}} }
\def\lsim{ \lower .75ex \hbox{$\sim$} \llap{\raise .27ex \hbox{$<$}} }

\title[Dark-matter fractions with IllustrisTNG]{The fraction of dark matter within galaxies from the IllustrisTNG simulations}
\author[M. R. Lovell et al.]{Mark R. Lovell$^{1,2,3}$\thanks{E-mail:lovell@hi.is}, Annalisa Pillepich$^{3}$, Shy Genel$^{4,5}$, Dylan Nelson$^{6}$, \newauthor  Volker Springel$^{7,8,6}$, R\"udiger Pakmor$^{7}$, Federico Marinacci$^{9}$, Rainer Weinberger$^{7}$, \newauthor Paul Torrey$^{9,10}$, Mark Vogelsberger$^{9}$\thanks{Alfred P. Sloan Fellow}, Adebusola Alabi$^{11}$ and Lars Hernquist$^{12}$ \\ 
$^{1}$Center for Astrophysics and Cosmology, Science Institute, University of Iceland, Dunhagi 5, 107 Reykjavik, Iceland \\
$^{2}$Institute for Computational Cosmology, Durham University, South Road, Durham DH1 3LE, UK\\
$^{3}$Max-Planck-Institut f\"ur Astronomie, K\"onigstuhl 17, D-69117 Heidelberg, Germany\\
$^{4}$Center for Computational Astrophysics, Flatiron Institute, 162 Fifth Avenue, New York, NY 10010, USA\\
$^{5}$Columbia Astrophysics Laboratory, Columbia University, 550 West 120th Street, New York, NY 10027, USA\\
$^{6}$Max-Planck-Institut f\"ur Astrophysik, Karl-Schwarzschild-Str. 1, 85741 Garching, Germany\\
$^{7}$Heidelberg Institute for Theoretical Studies, Schloss-Wolfsbrunnenweg 35, D-69118 Heidelberg, Germany \\
$^{8}$Zentrum f\"ur Astronomie der Universita ̈t Heidelberg, ARI, Mo ̈nchhofstr. 12-14, D-69120 Heidelberg, Germany\\
$^{9}$Department of Physics, Kavli Institute for Astrophysics and Space Research, Massachusetts Institute of Technology, Cambridge, MA 02139, USA\\
$^{10}$Department of Astronomy, University of Florida, 211 Bryant Space Sciences Center, Gainesville, FL 32611 USA\\
$^{11}$University of California Observatories, 1156 High St., Santa Cruz, CA 95064, USA\\
$^{12}$Harvard-Smithsonian Center for Astrophysics, 60 Garden Street, Cambridge, MA 02138, USA}

\date{Accepted *** Received ***; in original
form ***} 
\pubyear{2017}

\begin{document}

\label{firstpage}
\pagerange{\pageref{firstpage}--\pageref{lastpage}} 

\maketitle

\begin{abstract}
We use the IllustrisTNG (TNG) cosmological simulations to provide theoretical expectations for the dark matter mass fractions (DMFs) and circular velocity profiles of galaxies. TNG predicts flat circular velocity curves for $z = 0$ Milky Way (MW)-like galaxies beyond a few kpc from the galaxy centre, in better agreement with observational constraints than its predecessor, Illustris. TNG also predicts an enhancement of the dark matter mass within the 3D stellar half-mass radius ($r_\rmn{half}$; $M_\rmn{200c} = 10^{10}-10^{13}\Msun$, $z \le2$) compared to its dark matter only and Illustris counterparts. This enhancement leads TNG present-day galaxies to be dominated by  dark matter within their inner regions, with $f_\rmn{DM}(<r_\rmn{half})\gtrsim0.5$ at all masses and with a minimum for MW-mass galaxies. The 1$\sigma$ scatter is $\lesssim$ 10~per~cent at all apertures, which is smaller than that inferred by some observational datasets, e.g. 40 per cent from the SLUGGS survey. TNG agrees with the majority of the observationally inferred values for elliptical galaxies once a consistent IMF is adopted (Chabrier) and the DMFs are measured within the same apertures. The DMFs measured within $r_\rmn{half}$ increase towards lower redshifts: this evolution is dominated by the increase in galaxy size with time. At $z\sim2$, the DMF in disc-like TNG galaxies decreases with increasing galaxy mass, with $f_\rmn{DM}(< r_\rmn{half}) \sim 0.10-0.65$ for $10^{10} \lesssim M_{\rm stars}/\Msun \lesssim 10^{12}$, and are two times higher than if TNG galaxies resided in Navarro--Frenk--White dark matter haloes unaffected by baryonic physics. It remains to be properly assessed whether recent observational estimates of the DMFs at $z\sim2$ rule out the contraction of the dark matter haloes predicted by the TNG model. 

\end{abstract}


  \section{Introduction}
  \label{intro}

   It has been known for many years that galaxies contain a substantial fraction of their mass in non-luminous dark matter in addition to the visible contributions from stars and gas. The quantity of dark matter in the Universe at large has been well established, e.g. by cosmic-microwave background experiments \citep{wmap11,PlanckCP15}. Hence, the focus of the field of dark matter astrophysics has now shifted to the more difficult problem of understanding how dark matter condenses in concert with the baryonic matter within the high density environments of galaxies. The extension of precision science to this regime has been made possible by the introduction of new models of galaxy formation, backed up by powerful supercomputing facilities, in order to make theoretical predictions, and by a range of new observational techniques that probe the distribution of the various galactic matter components: microlensing techniques in our own Galaxy \citep[e.g.][]{Wegg16}, Doppler shifts in the spectra of stellar and gaseous components in other galaxies \citep[e.g.][]{Bekeraite16}, as well as  dynamical modelling based on their satellite and globular cluster populations \citep{Alabi17}, as we describe below. 
   
   The development of multi-slit spectrographs and integral field units (IFU) has enabled researchers to map the kinematics of galactic matter components from the inner parts of galaxies \citep[SINS survey;][]{ForsterSchreiber09} out to many effective radii (\citealp[e.g.][]{vanDokkum15}; \citealp[SAMI survey:][]{Cecil16}; \citealp[CALIFA survey:][]{ZhuL18}). Recently, the {\sc sluggs} survey \citep{Brodie14} has combined Keck/DEIMOS and Subaru/SuprimeCam measurements of globular clusters to infer the mass fraction of dark matter to total within elliptical galaxies in the low-redshift Universe, out to five effective radii. A wide range of different dark matter mass fractions within this aperture have been determined, from $\sim0.1$ to 0.9 \citep{Alabi16,Alabi17}. In fact, the innermost regions of low-redshift early-type galaxies had been probed previously by using similar methods with the {\sc sauron} and {\sc atlas$^\rmn{3D}$} surveys \citep{Emsellem04,Cappellari13}. They have successfully detected the presence of dark matter even in these regimes where baryon densities are very high, while at the same time finding that dark matter plays a very small dynamical role within one effective radius, with dark matter mass fractions almost always lower than 20~per~cent \citep{Cappellari13}. 
   
   Further constraints on the total masses of galaxies have been obtained from a series of HI surveys, such as ALFALFA and HIPASS, which have shown that the relationship between a galaxy's total mass and its stellar content is non-trivial \citep{Papastergis2011,Bekeraite16}. Dynamical models applied to optical and infra-red spectra and photometry have provided a third set of measurements of the dark matter fraction, often in conjunction with other methods \citep{Thomas07,Tortora09,Humphrey10,Toft12,Tortora12,Beifiori14, Tortora14,Tortora16a,Tortora16b,Oldham18}. Finally, an extra set of constraints have been obtained from lensing studies, which have the benefit of probing the 3D half-stellar mass rather 2D half-light radii and are therefore easier to compare to simulations \citep{Koopmans06,Tortora10,Barnabe11,JimenezVicente15}. These studies report a wide diversity of results concerning the dark matter content of early type galaxies (ETGs) in particular, including its amplitude and redshift evolution. They also predict very different stellar mass contributions, and this is something that we will explore in our comparison to observations.

In order to be complete,  theories of galaxy formation must accurately predict the dark matter content of all galaxies simultaneously, including those of different morphologies and at higher redshifts. Recently, \mbox{\citet{Genzel17}} have obtained rotation curves of $z=2$ star-forming galaxies using SINFONI and KMOS measurements of rest-frame H$\alpha$ emission. These rotation curves are falling in amplitude as a function of radius \citep{Lang2017}, and have dark matter fractions $\sim0.4$~dex lower than disc galaxies at $z=0$ \citep[e.g.][]{Courteau15}. That many of these observed galaxies exhibit behaviours different from the flat rotation curve of local disc-galaxies \citep{Rubin80}, including our own \citep{Bovy13,IPB15}, suggests that galactic structure is rich and diverse, and a challenge for state-of-the-art theories of galaxy formation: the Magneticum simulations have already shown promise in tackling these issues effectively \citep{Teklu17}. Simultaneously, the inferred dark matter fractions in our own Galaxy vary in the $0.3-0.6$ range according to different analyses and methodologies \citep[][for $6 - 10$~kpc apertures]{Bovy13,IPB15,BlandHawthorn16,Wegg16}: such discrepancies are a reminder of how complex observational analyses and dynamical models can be for the purposes of determining the matter balance within galaxies. 
  
From a theoretical perspective, the standard paradigm proposes that galaxies form and evolve in virialized haloes of cold dark matter (CDM).
DM is widely expected to form a collisionless \citep[but see e.g.][ for a discussion of collisional models]{Vogelsberger12,Vogelsberger14c} perturbed fluid that collapses under gravity. The evolution of the DM fluid is impossible to predict analytically due to the highly non-linear nature of the equations at redshifts below $z~\lsim~50$, and is therefore modelled using numerical simulations, e.g. dark matter-only (DMO) cosmological simulations that treat 100~per~cent of the matter in the Universe as collisionless dark matter.
  CDM haloes simulated with DMO numerical models have density profiles that are cuspy rather than cored in their centres, and are well described by the Navarro--Frenk--White \citep[NFW;][]{NFW_96,NFW_97} or Einasto \citep{Einasto65} density profiles. For example, the NFW profile corresponds to a dark matter circular velocity curve that rises within the inner 15~kpc of those haloes sufficiently massive \mbox{\citep[$>10^{8}\Msun$,][]{Benson_02}} to form galaxies, and in this mass range is distributed in a smooth component rather than in lumpy subhaloes \citep{Springel08b,Vogelsberger09}. This increase in amplitude of the DM circular velocity curve explains the flatness of observed galaxy rotation curves: the centripetal acceleration due to dark matter increases at the same rate as the acceleration due to the stellar component of the galaxy, which is much more concentrated than the dark matter, decreases.
 
  Recent advances in galaxy formation models and the availability of computing power have enabled astronomers to go beyond the DMO approach, by including gas physics and many complex astrophysics processes in simulations of structure formation \citep[e.g.][]{Springel03,Springel05b,Schaye10,Crain10,Hopkins14,Vogelsberger14b,Schaye15,Somerville15,WangL15,Beck16,Hopkins18}. In addition to solving for the hydrodynamical properties of the gas, galaxy formation simulations apply subgrid models to predict the formation of stars, the growth of black holes, and the energy injection from supernova explosions and active galactic nuclei (AGN) to star-forming and intra-cluster gas. Simulation efforts such as the EAGLE \citep{Schaye15}, Illustris \citep{Vogelsberger14b}, HorizonAGN \citep{Dubois14,Kaviraj17} and Magneticum \citep{Steinborn15} projects have enjoyed success in matching e.g. the observed galaxy luminosity function across cosmic times together with a plethora of other stellar- and gas-related galaxy properties: they are therefore able to make credible predictions for the effects of galaxy formation processes on the distribution of DM. 
  
  One weakness of these numerical experiments lies in the need for subgrid prescriptions, as they cannot be directly derived from models of e.g. individual star formation {\it ab initio} and must therefore be constructed from a range of assumptions. The expense involved in running simulations with sufficient resolution and volume also introduces challenges for accurately calibrating the models against observations. Therefore, the ensemble of the adopted construction hypotheses can lead to rather different expectations across different simulation codes and galaxy formation models in relation to e.g. structural galaxy properties such as the circular velocity profiles of galaxies \citep{Scannapieco12} or the gas mass content within haloes \citep[see e.g.][for a discussion]{Pillepich17b}. For example, simulations like EAGLE \citep{Schaller15}, NIHAO \citep{WangL15}, Auriga \citep{Grand17}, and Magneticum \citep{Teklu17} predict Milky Way-analogue (MW-analogue) circular velocity curves that are flat outside a few kpc from the centre and in qualitatively good agreement with observations, whereas Illustris \citep{Genel14, Vogelsberger14d} instead predicts circular velocity curves that are rising below $\sim15$~kpc and the FIRE-2 project galaxies \citep{GarrisonKimmel17} have circular velocity curves that instead slightly decrease in amplitude in the same radius range.

  In this study we consider a new set of simulations that are part of the IllustrisTNG project \citep{Nelson17, Naiman18,Springel17,Pillepich17b,Marinacci17}. IllustrisTNG builds upon the technical and scientific achievements of its predecessor, the Illustris simulation (\citealt{Vogelsberger14d}; \citealt{Vogelsberger14b}; \citealt{Genel14}; \mbox{\citealt{Sijacki15}}). It both upgrades the scope of Illustris as well as introduces new physics, of which a BH-driven wind and cosmological magnetic fields are prime examples. We use the IllustrisTNG simulations to predict the mass fraction of dark matter within galaxies, as an emergent feature of an effective model that has been tailored to reproduce basic galaxy population statistics \citep[see ][for an overview and Section \ref{sims}]{Pillepich17,Weinberger17}.
   We take advantage of the large dynamic range of galaxies produced in the IllustrisTNG simulations, from brightest cluster galaxies to massive dwarfs -- a successful model of galaxy formation should predict accurately as many of the available observables at all mass scales simultaneously as possible. In particular, we measure the matter distributions in spherical symmetry, circular velocity profiles, the dark matter fraction profiles, and their values within fixed galactocentric distances, for a first-order comparison to the estimates inferred from observations. We build an understanding about the underlying effects by comparing the matter distribution within IllustrisTNG haloes to those obtained in their dark matter-only, gravity only, counterparts and, when possible and relevant, to those modelled in Illustris.

   This paper is organised as follows. In Section~\ref{sims} we present a summary of the simulations used in this work. Along with the method for identifying dark matter haloes and the properties of their galaxies, we present our definitions, galaxy samples and observables in Section~\ref{meth}. Our results, which entail a quantification of the matter distribution within simulated haloes,  comparisons between simulation approaches, and effects of baryonic physics onto the phase-space properties of dark matter, are given in Section~\ref{res} along with a discussion in the context of galaxy formation theory. We discuss these results in comparisons to observations in Section~\ref{dis} and we draw conclusions in Section~\ref{con}.

  \begin{table*} 
\begin{center}
 \caption{Table of the simulation properties used in this work. These include the size of the cubic periodic volume, the number of dark matter particles, which is also equal to the number of initial gas cells; the dark matter particle mass $m_{DM}$; the dark matter softening length, which is physical for $z\le1$, comoving for $z>1$, and equals the gravitational softening of the stellar particles for TNG (but not Illustris, in which the stellar particle softening is half that of the dark matter particles at $z=0$ and evolves differently for $0<z<1$); the cosmology and the galaxy formation model. See \url{http://www.tng-project.org/} and references therein for a full list of simulation parameters. }

\begin{tabular}{lcccccc}
\hline
Name &  Box Length &\#DM elements & $m_\rmn{DM}$ & $\epsilon_\rmn{DM}$ & Cosmology & Galaxy Model \\
 &  [Mpc] & & [$\Msun$] & [kpc] & &  \\

\hline
TNG100 & 106.5 & $1820^3$ &$7.5\times10^6$ & 0.74 & Planck & IllustrisTNG \\
TNG100-DM &  106.5 & $1820^3$& $8.9\times10^6$ & 0.74 & Planck & DMO \\
\hline
TNG300 & 302.6 & $2500^3$ & $5.9\times10^7$ & 1.48 & Planck & IllustrisTNG \\
TNG300-DM  & 302.6 & $2500^3$ & $7.0\times10^7$ & 1.48 & Planck & DMO \\
\hline
Illustris & 106.5& $1820^3$ & $7.5\times10^6$ & 1.42 & WMAP7 & Illustris \\
Illustris-Dark & 106.5& $1820^3$ & $8.9\times10^6$ & 1.42 & WMAP7 & DMO \\
\hline
\end{tabular}
\end{center}
\label{simstab}
\end{table*}

   \section{Galaxy formation model and simulations}
   \label{sims}
   
   We make predictions for the properties of galaxies by using the IllustrisTNG model (hereafter TNG for brevity), which is fully described in \cite{Weinberger17} and \cite{Pillepich17} and is optimized to simulate galaxy populations in large cosmological volumes. 
   The TNG model features gas cooling, star-formation, stellar evolution and enrichment, galactic winds \citep{Pillepich17}, feedback from active galactic nuclei \citep[AGN,][]{Weinberger17}, and magnetic fields \citep{Pakmor11}. It is an upgrade of the Illustris model \mbox{\citep{Vogelsberger13a, Torrey15}}, and was designed to mitigate the shortcomings of the Illustris simulation \mbox{\citep{Vogelsberger14b,Genel14,Sijacki15}}, whilst gaining insights into how the required changes to the subgrid implementation reflect how the underlying physical processes are different to what was assumed when developing the previous model as summarised in Section~6 of \citet{Nelson15}.
   
 The IllustrisTNG model is implemented in the eponymous IllustrisTNG simulation project\footnote{\url{www.tng-project.org}}. The suite comprises two currently completed main simulations, dubbed TNG100 and TNG300 \citep{Marinacci17, Nelson17, Pillepich17b, Springel17,Naiman18}. They both use the hydrodynamical code {\sc arepo} \citep{Springel10}  in its galaxy formation mode, as did Illustris. TNG100 is a cosmological box of side length 75~$h^{-1}$Mpc$~\approx100$~Mpc, and is a direct counterpart to the original Illustris simulation. It has the same initial number of resolution elements, $2\times1820^{3}$, and the same softening lengths for the star particles, although the dark matter softening lengths have been halved in size for TNG relative to Illustris at $z=0$ and evolve differently for $0<z<1$. However, the linear matter power spectrum and cosmological parameters applied are those of the Planck cosmology \mbox{\citep{PlanckCP13,PlanckCP15}} rather than the WMAP-7 cosmology \mbox{\citep{wmap11}} used in Illustris.  
 The TNG300 simulation comprises a volume over 20 times that of TNG100 (side length of 205~$h^{-1}$Mpc~$\approx300$~Mpc, and $2\times2500^{3}$ resolution elements) but performed at a factor 8 (2) lower mass (spatial) resolution. It therefore offers excellent statistics on clusters of galaxies, and we also use it as a way to assess resolution convergence. 
 
 All TNG simulations were performed with the same galaxy formation model and, crucially, the same values of the model parameters but for the particle softening lengths and the black hole kernel-weighted neighbour number \citep[see][]{Pillepich17}. 
 Unlike the TNG100 case, there are no Illustris counterparts to the TNG300 run. 
 All TNG runs are paired with a DMO counterpart with the same initial condition phases and in which all of the matter is treated as cold, collisionless dark matter particles. These are labelled as e.g. TNG100-DM and TNG300-DM.  Note that TNG100-DM is identical to Illustris-Dark but for the change in cosmology and the $z<1$ softening lengths. We summarise the important statistics of the simulations featuring in this paper in Table~\ref{simstab}. 

The TNG model and simulations have been designed with the goal of reproducing simulated galaxy and halo populations in reasonable agreement with the observed ones, namely: the global star formation rate density evolution, the $z=0$ stellar-mass to halo-mass relation, the $z=0$ galaxy stellar mass function, the $z=0$ BH mass vs. galaxy stellar mass relation, the average $z=0$ gas fraction within the virial radius as a function of halo mass, and the average $z=0$ galaxy sizes, all of which are collectively discussed in comparison to the Illustris results by \cite{Pillepich17} on a series of smaller test simulated volumes. In fact, the TNG100 and TNG300 outcome is broadly consistent with a series of observations, including the galaxy stellar mass functions at low redshifts \citep{Pillepich17b}, the large-scale spatial clustering of galaxies \citep{Springel17}, the low-metallicity spread of stellar Europium in MW-like galaxies \citep{Naiman18}, the metallicity content of the intra-cluster medium \mbox{\citep{Vogelsberger17}}, and the gas-phase oxygen abundance and distribution within \citep{Torrey17} and around galaxies \citep{Nelson17b}. Careful quantitative comparisons to observed data have shown that TNG galaxies reproduce to an  excellent degree of agreement the galaxy colour bimodality observed in the Sloan Digital Sky Survey \mbox{\citep{Nelson17}} and the average trends, evolution, and scatter of the galaxy size-mass relation at $z\lesssim2$ \citep{Genel17}. These model validations support the usage of the TNG galaxy population as a plausible synthetic dataset, with which to further assess how the different matter components balance each other within the inner regions of haloes where galaxies reside.

   \section{Methods and sample selection}
   \label{meth}
   
In this section we describe our halo / galaxy selections, our choice of observables and how we calculate these observables in the context of simulations in comparison to that done in observations.

\subsection{Matching Algorithm}
\label{sec:matching}
In order to assess the precise effect of the baryonic physics model on the halo with respect to previous models -- such as DMO and the original Illustris simulations -- we apply matching algorithms to identify pairs of {\it analogue} haloes across simulations. Objects that originated from the same Lagrangian patch of the initial conditions may be considered `analogues', i.e. the `same' halo. We proceed in two complementary ways,  since some datasets have only one type of matching catalogue available: 1) we determine DMO counterparts to TNG haloes by using a method based on the IDs of the halo member particles, where two counterpart haloes that contain most of each others' particles are considered a match; and 2) we determine Illustris counterparts to TNG haloes using the Lagrangian region matching algorithm of \citet{Lovell14}, the latter of which we summarise below.

We define our `haloes' as a self-bound collection of dark matter particles, as identified by {\sc subfind} (see Section.~\ref{sec:sel} for further details). The initial conditions for a Lagrangian patch of each halo is determined using the dark matter particles that the halo has at the snapshot at which it achieves its maximum mass (but see below for other snapshot choices), and the continuous nature of the matter distribution is achieved by modelling each particle as a spherical shell of radius equal to the interparticle separation. The Lagrangian patch is then compared to the patches of haloes in the companion simulation by means of their density distributions and gravitational potentials to obtain a quality-of-match statistic, $R$, that has the value 1 for a perfect match and $<1$ otherwise; see equation~(6) and the accompanying discussion of \citet{Lovell14} for a comprehensive presentation. The companion simulation halo with the highest value of $R$ is then considered the `match'. We have found that the highest quality match typically has a value $R>0.95$ and the second-highest (and therefore not a match) $R<0.5$.

This procedure is computationally expensive, therefore we take the following steps to make the algorithm more efficient. For haloes with more than 1000 particles, we select a maximum of 1000 particles at random to describe the Lagrangian patch and increase the effective shell radius to match the new, higher effective interparticle separation of the sample. We attempt to match pairs of haloes that are within a factor of two in mass and whose Lagrangian patch centres-of-mass are within 2.5~Mpc (12.5~Mpc) of each other for haloes of mass $<10^{12}\Msun$ ($\ge10^{12}\Msun$). Finally, we note that our algorithm may be compromised if a merger is ongoing in one simulation and not in the second, including at the maximum mass snapshot, due to the manner in which {\sc subfind} identifies haloes. We therefore mitigate this problem by repeating the process with particles taken from the half-maximum mass snapshot; whichever snapshot returns the higher value of $R$ then determines the matched halo.

This method enables us to match between multiple classes of virialized objects and between any two simulations that use the same initial conditions volume. In this study it is used to identify Illustris counterparts to TNG100 central haloes; in future papers it will be employed to match satellite galaxies, to match galaxies between resolutions, and also between different models.

  \begin{figure}
         \includegraphics[scale=0.44]{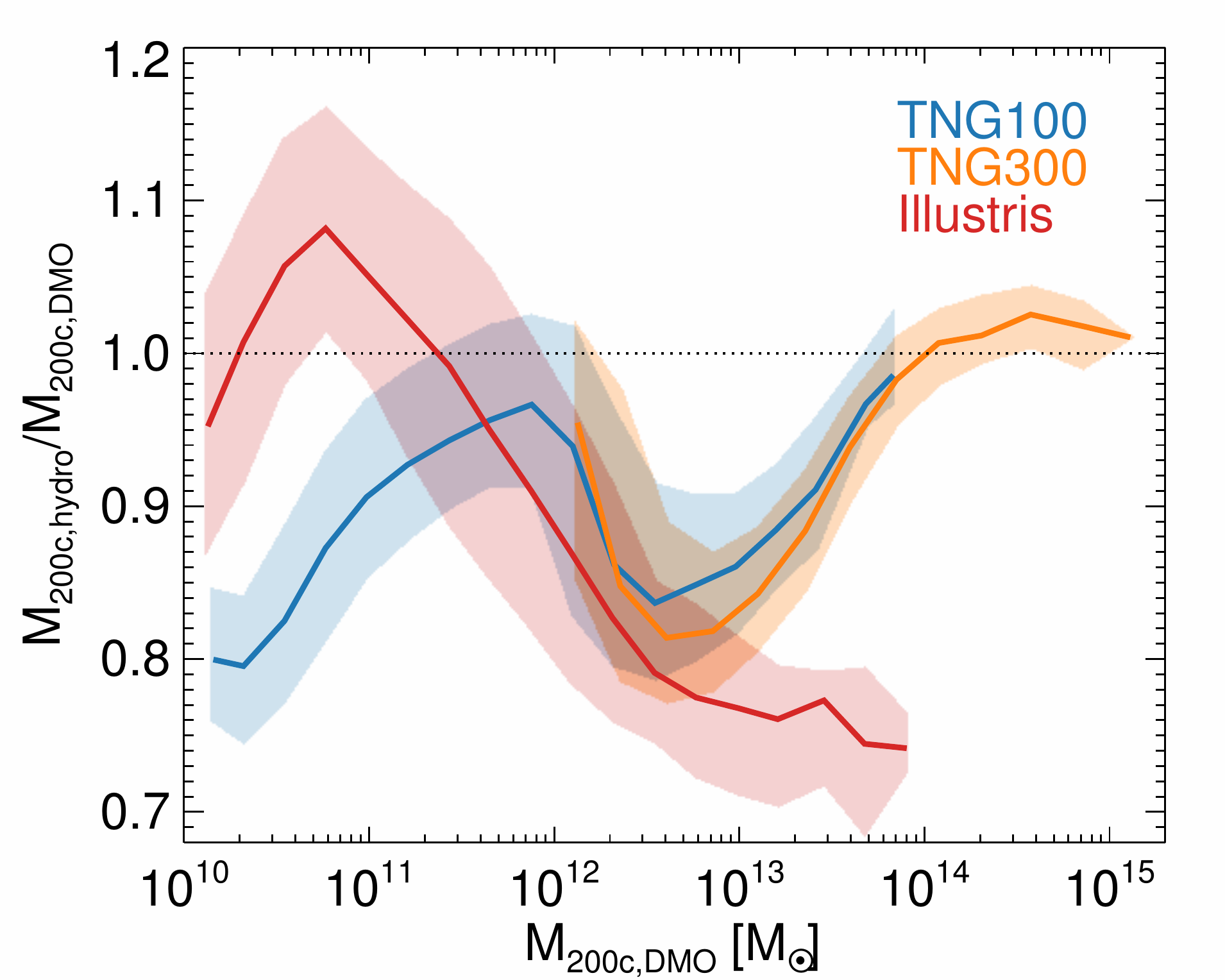}
        \caption{The ratio of hydro-to-DM-only $ M_{200c}$ for TNG100 haloes matched to TNG100-DM counterparts (blue data), Illustris-1 to Illustris-1-Dark (red data) and likewise for TNG300 haloes (orange). The shaded regions enclose 68~per~cent of the data. The bin centres are taken to be the median $M_\rmn{200c,DMO}$ of the set of binned galaxies.}
        \label{M200Func}
   \end{figure}
   
\subsection{Galaxy selection and halo mass}
\label{sec:sel}
Haloes and subhaloes are identified using the {\sc subfind} subhalo finder code \citep{Springel01}. Isolated haloes are detected by means of a geometric friends-of-friends (FoF) algorithm, and are then separated out into self-bound haloes and subhaloes using a gravitational unbinding procedure. The central halo is the largest self-bound structure within the parent FoF halo, the smaller structures are identified as subhaloes. By association, the baryonic galaxy hosted by the central halo is considered the `central galaxy', and those galaxies hosted within subhaloes are labelled satellite galaxies.

In this paper we consider only {\it central} galaxies. We classify our galaxies primarily by means of the total mass of their host halo: throughout, we use $M_{200c}$, which is the total mass enclosed within the radius that encompasses a mean density 200 times that of the critical density of the Universe.
In our analysis we focus on the halo mass range $M_{200c}=10^{10}-10^{15}~\Msun$, with the TNG300 volume providing the statistics needed to study haloes above $10^{14}~\Msun$ and hence the massive galaxies at their centres. 

The values of $M_{200c}$ are not insensitive to the adopted galaxy physics model. This may be surprising, given that the total halo mass extends to much larger galactocentric distances than the baryonic material where physical mechanisms other than gravity dominate. Nevertheless, this has been already demonstrated in a number of studies across models \citep{Vogelsberger14d, Schaller15, Chua16}. We further quantify this assertion in the TNG simulations by comparing the value of $M_{200c}$ for each of our galaxies to their counterparts in the DMO simulation run of the same volume. The halo by halo matching is performed with the algorithm introduced in Section \ref{sec:matching}. We compute the ratio of the two $M_{200c}$ values -- the hydrodynamical or full-physics run to the DMO run -- and plot median results and 1-sigma regions as a function of the DMO $M_{\rm 200c, DMO}$ in Fig.~\ref{M200Func}; we include haloes from the TNG100 (blue), TNG300 (orange) and Illustris-1 (red) runs. 

In both the Illustris and TNG models there is a significant deviation from the assumption that $M_{200c}$ is unaffected by hydrodynamical/galaxy physics processes. Halo masses $>10^{12}\Msun$ are suppressed in both models, by a median of up to 15~per~cent in TNG100 and 23~per~cent in Illustris-1. In Illustris it is the most massive haloes in the sample that experience the strongest suppression, whereas in TNG the decrease is strongest at $3\times10^{12}\Msun$ before returning almost to equality with DMO at the highest masses. The TNG300 run shows little change in halo mass with the inclusion of baryon physics for $M_{200c}>10^{14}\Msun$, and if anything $M_\rmn{200c}$ values that are a few per~cent higher in the full physics run. The TNG300 run shows the same behaviour as TNG100 within ~5~per~cent in the overlapping mass range, confirming good convergence in this observable.  Below $10^{12}\Msun$ the Illustris and TNG models behave differently, with the masses of $10^{11}\Msun$ Illustris haloes being {\it higher} compared to the DMO expectation rather than suppressed as in TNG.  In this regard, the TNG100 simulation produces similar results to the {\sc eagle} and {\sc apostle} projects \mbox{\citep{Schaye15,Fattahi16,Sawala16a}} in which the halo mass is progressively suppressed for lower mass haloes. However, those models predict a gradual rise in the ratio value, from $\sim0.7$ at $10^{10}$ to $1.0$ at $5\times10^{13}\Msun$ rather than an abrupt increase to almost unity at $8\times10^{12}\Msun$ followed by a trough; the measurement of this ratio is therefore an interesting tool to understand the behaviour of different feedback models. 
Similar results from the TNG simulations have been shown already by \cite{Springel17}, although we adopt a different method to ``match'' haloes across simulations. Fig.~\ref{M200Func} simultaneously verifies the functioning of our matching algorithm and confirms the quantitative findings of \cite{Springel17}, namely that the value of $M_{200c}$ rarely changes by more than 30~per~cent relative to the DMO expectation with a magnitude that may vary across models. 

\subsection{3D galaxy sizes, stellar masses, and morphologies}

Galaxies are further characterized by their size, stellar mass and morphology.

A careful study of galaxy sizes in the TNG simulations has been performed in \citet{Genel17}; here we present a summary of the methods and results relevant for this paper. We parametrize our sizes using the 3D stellar half-mass radius, $r_\rmn{half}$\footnote{We always use this symbol to refer to the 3D stellar half-mass radius, and never for the 2D projected version. Note that even though the half-light radius is a projected quantity, the mass enclosed within it is not unless stated otherwise.}, defined as the radius that encloses half of the bound stellar mass of the galaxy as determined by the {\sc subfind} algorithm. Strictly speaking, a true half-light radius requires that we compute the light profile in some photometry band and take account of the difference between 2D projected and 3D de-projected radii; the differences between 3D stellar mass sizes and 2D stellar light sizes can be found in  \citet{Genel17}, with the former being up to 1.6 times larger than the latter and the two being equivalent for galaxies with stellar mass larger than about $10^{10.5}\Msun$.

The stellar half-mass radius also gives us a way to define the stellar masses of galaxies. The bound stellar mass as found by {\sc subfind} does not have a good observational comparison, and therefore another method should be identified, particularly for brightest cluster galaxies (BCGs) in which it can be difficult to determine where the galaxy ends and the intra-cluster light begins \citep[see e.g.][]{Pillepich17b}. Throughout, we therefore adopt our stellar masses to be the mass in star particles enclosed within $2\times r_\rmn{half}$, and label this mass $M_{*}$. The relationship between stellar mass and halo mass is by design close but not identical to the expectations of the empirical models of \citet[e.g.][]{Behroozi13,Moster13}; it is available as Fig.~11 of \citet{Pillepich17b}.

As shown in \citet{Genel17} and \citet{Pillepich17}, TNG and Illustris make similar predictions for the sizes of galaxies with stellar masses above $10^{11}\Msun$, the median expectation increasing from 10~kpc at $10^{11}\Msun$ to $\sim22$~kpc at $6\times10^{11}\Msun$. However, the models diverge substantially at lower stellar masses, with Illustris galaxies being consistently twice  the size of their TNG counterparts at fixed stellar mass for $M_{*}<10^{10}\Msun$\footnote{ The difference between the galaxy radii in Illustris and TNG is highly non-trivial: see \citet{Pillepich17} for a discussion on this topic.} and TNG galaxies being in much better agreement with observational constraints \citep{Genel17}. This result will be crucial when we consider the dark matter fractions within galaxies. 

One of the possible observational methods to separate galaxies by morphology consists in fitting light profiles. Here we instead take advantage of the correlation between morphological type and the galaxy kinematics to divide our sample into {\it disc-dominated} and {\it bulge-dominated} galaxies. The kinematics are described by the circularity distributions \citep{Scannapieco12} and are calculated for the TNG simulations following equation~(1) of \mbox{\cite{Marinacci14}} and using all of the star particles within 10 stellar half-mass radii. Star particles rotating in the same sense as the galaxy as a whole have a circularity of 1, those rotating in the opposite sense score -1, and other bound orbits an intermediate value. We define our bulge/early-type component as the total stellar mass that exists below 0 multiplied by two (assuming a symmetric distribution of bulge particle orbits) and the disc/late-type component as the total stellar mass minus the bulge mass. We label a galaxy as bulge-dominated if more than 50~per~cent of the mass is in the bulge component, and otherwise as disc-dominated; note that we always label our galaxies based on the circularities calculated within 2 half-stellar mass radii for TNG galaxies\footnote{For Illustris we use 10 half-stellar mass radii, as the 2 half-stellar mass radii measurements are not available. The choice of 10 radii in the older simulation follows from the work of \citet{Romanowsky12}.}.

\begin{figure*}
    	\includegraphics[scale=1.46]{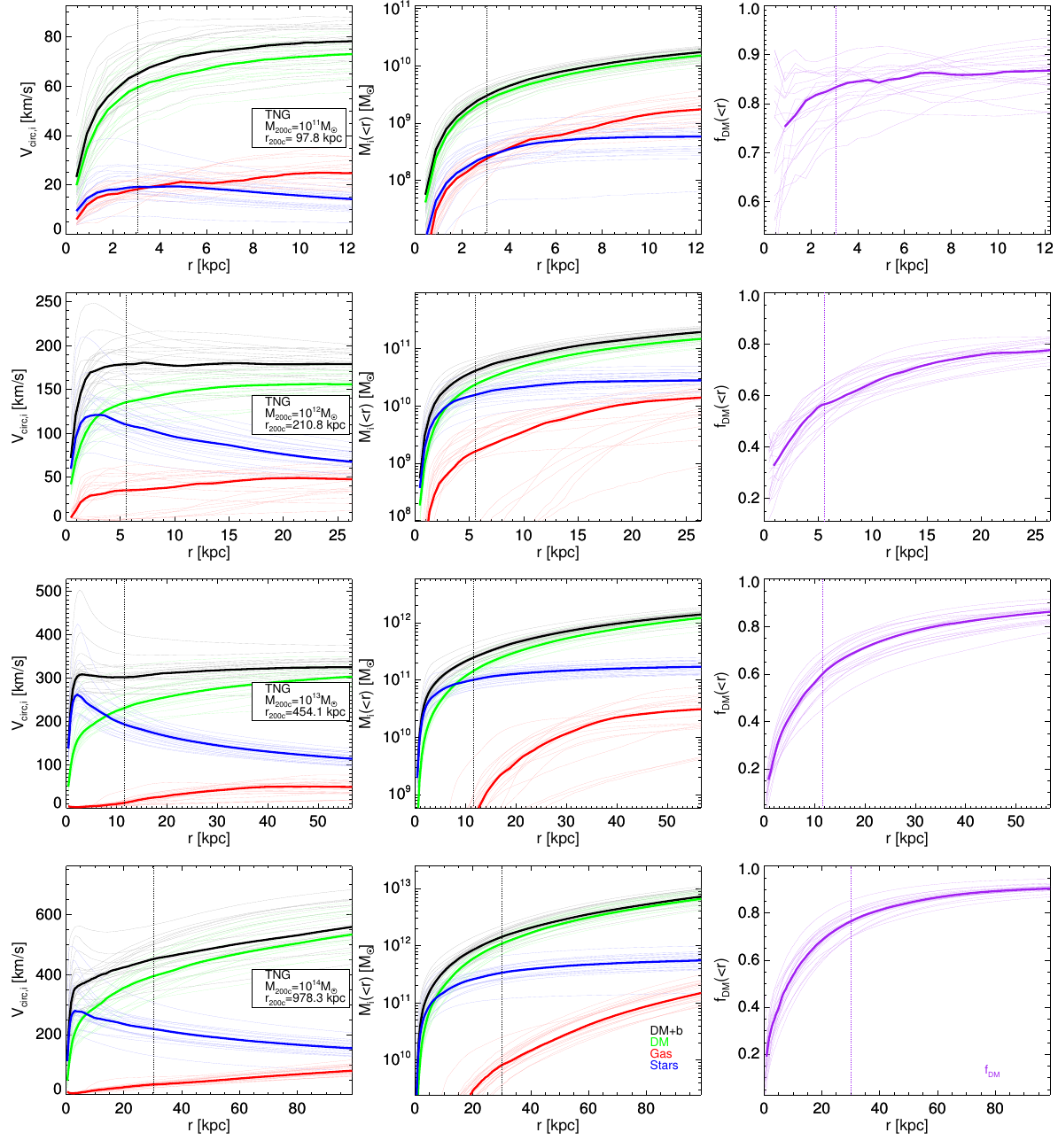}
        \caption{$z=0$ circular velocity profiles for a sample of TNG central galaxies (left-hand panels), the corresponding enclosed mass profiles (central panels) and dark matter fraction profiles (right-hand panels). Each row corresponds to a different host halo mass, starting from top: $M_{200c}=10^{11}, 10^{12},10^{13},10^{14}\Msun$. Twenty individual haloes are shown with thin curves, and their median profile is shown with a thick curve. The average value of $r_\rmn{half}$ is shown as a vertical dotted line. The $10^{14}\Msun$ galaxies are drawn from the TNG300 simulation, and the other three sets from TNG100. Colours correspond to different mass components, and are given in the panel legends.}
        \label{VcircM}
\end{figure*}

 \subsection{Observables: matter profiles, circular velocity curves, dark matter concentrations, and dark matter fractions}

Having made our galaxy selection, our primary goal is to measure their mass distributions as a function of galactocentric distance and compare these to observational constraints. We quantify the mass enclosed profiles within galaxies by matter component (stars, gas, dark matter and the sum of all three -- denoted as ``DM+b''). The profiles are measured in linear bins from a minimum galactocentric distance of 0~kpc to ten percent of the virial radius ($R_{200c}$). Such an outer boundary corresponds to about five times the stellar half mass radius for the majority of the analyzed galaxies, placing our focus at the centre of the dark matter haloes.

We mimic the observed galaxy rotation curves by measuring the circular velocity curve for our simulated galaxies, $V_{\rmn{c}}(r)$. The latter is defined as the velocity of a test particle due to its centripetal acceleration at a distance $r$ from the centre of the matter distribution of enclosed mass $M(<r)$, and is given by the equation:

 \begin{equation}
 	V_{\rmn{c}}(<r) = \sqrt{\frac{GM(<r)}{r}},
 \end{equation}

 \noindent
 where $G$ is the universal gravitational constant. This is the velocity that the test particle would have if moving in a circular orbit, and should therefore be a good approximation for the net rotation of disc galaxies. This quantity is simple to calculate using simulation data, can be measured for any galaxy type and separately for the dark matter (labelled DM), the two baryonic components (gas and stars) and the three components combined (DM+b). However, the theoretical circular velocity curve is not strictly the same as the rotation curves probed observationally.
 
We use \vmax to denote the peak of the circular velocity curve: in DMO frameworks, this is often considered a proxy for the total halo mass. \rmax is the radius at which \vmax occurs. Both quantities can be identified for the total mass curve or by component. At fixed \vmax, smaller \rmax values imply higher galaxy or halo concentrations. As a simple measure for the dark matter halo concentration, we will use the ratio from equation~(6) of \cite{Springel08b}, namely $\delta V \propto ($\vmax / \rmax$)^2$, where all quantities are to be intended for the dark matter component alone.
 
Finally, we extract from the simulated galaxies the dark matter mass fraction profiles ($f_\rmn{DM}(<r)$), i.e. the profiles of the ratio between dark matter enclosed mass and total mass,  and tabulate for every selected galaxy the dark-matter fractions (DMFs) within specific fixed apertures: e.g. five physical kpc, one times the stellar half-mass radius and five times the stellar half-mass radius. This is done in analogy to that done in observations  e.g. one stellar mass radius for \citet{Cappellari13} and five radii for \citet{Alabi17} and to probe both the innermost galaxy regions and their outskirts.

   \section{Results}
   \label{res}
   
   \subsection{Matter distribution within $z=0$ TNG galaxies}
      \label{res1}
      
   \subsubsection{Matter component profiles}

 We begin our presentation of the results with a detailed discussion of the matter distribution in a sample of haloes that span the available halo mass range. We generate four subsamples of haloes, each of which is centred on a halo mass, $M_\rmn{200c}\rmn{(sample)}=1\times10^{11}, 1\times10^{12},1\times10^{13}, 1\times10^{14}\Msun$, and consists of the 20 haloes closest in mass to each sample mass, in the fashion $M_\rmn{200c}(1)<M_\rmn{200c}(2)<...<M_\rmn{200c}(10)<M_\rmn{200c}\rmn{(sample)}<M_\rmn{200c}(11)<...<M_\rmn{200c}(20)$. The $1\times10^{14}\Msun$ sample is drawn from the TNG300 simulation and the other three from TNG100. 

Figure~\ref{VcircM} quantifies the spherically-symmetric distribution of all matter components both separately (gas, stars, dark matter) and combined (DM+b) within our four samples of galaxies, via the circular velocity curves (left-hand panels), the enclosed mass profiles (middle) and the dark matter mass fraction profiles (right). As outlined in the previous Section, we focus on profiles within 10~per~cent of the virial radius (whose averages are indicated in the insets), which corresponds to about 5 times the 3D stellar half mass radius ($r_\rmn{half}$) of the galaxies (whose averages are denoted by vertical dashed lines in each panel). In this Sub-section we limit the analysis to $z=0$ haloes.
 
At all halo masses $\ge10^{12}\Msun$, the gas component is a sub-dominant contribution to the total mass within many multiples of $r_\rmn{half}$, yet it exhibits the largest relative galaxy-to-galaxy variations and matches the stellar mass contribution only in haloes of $\sim10^{11}\Msun$.

\begin{figure}       
        \includegraphics[scale=0.65]{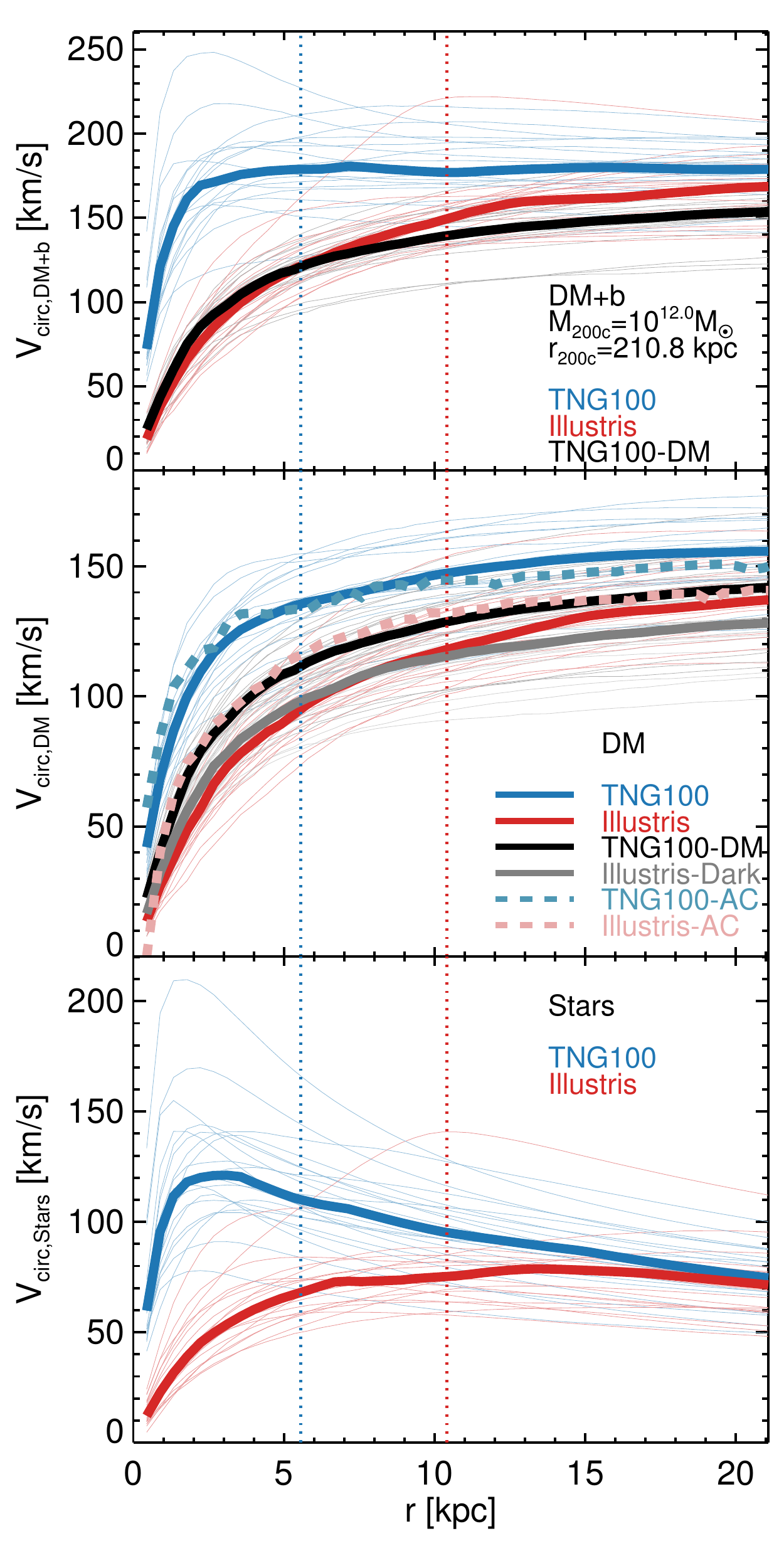}
        \caption{$z=0$ circular velocity profiles for the total (top row), dark matter (middle row), and stellar (bottom row) components of a sample of 20 TNG100 central galaxies with halo mass of about $10^{12}\Msun$ (blue curves), plus their matched Illustris counterparts (red). In the top two panels, we include the matched TNG100-DM counterparts (black). We also include the dark matter-only counterparts of the plotted Illustris galaxies, namely Illustris-Dark analogues (in grey) to the red curves (middle panel only). The differences between the dark-matter only sets (black vs. grey) arise from two factors: TNG100(-DM) and Illustris(-Dark) are performed with different cosmological parameters and the effective median mass of the analogue samples are slightly shifted according to the findings of Fig.~\ref{M200Func}. Finally, we include the average dark matter profile expected from adiabatic contraction \citep{Blumenthal86,Gnedin04} for TNG100 (Illustris) as a thick dashed light blue (pink) line in the middle panel.}
        \label{VcircByType}
\end{figure}

Apart from the gaseous component, the circular velocity profiles vary with halo mass. The $10^{11}\Msun$ haloes have rising (DM+b) profiles within $r_\rmn{half}$, as do the much more massive $10^{14}\Msun$ haloes to a lesser degree. The intermediate mass haloes exhibit a variety of different DM+b circular velocity curves within fractions of the $r_\rmn{half}$, all of which are flat between 5~kpc and the remainder of the visible galaxy (20-30~kpc): this is due to the larger proportion of mass in stars within $r_\rmn{half}$ as shown explicitly in the dark matter fraction curves. Many also show sharp peaks within 5~kpc. These peaks -- where we define a peaked galaxy generously (see Section~\ref{sec:Ddgat} for an alternative, observation-specific discussion of a peak) as a galaxy in which the circular velocity at 5~kpc is higher than that at 10~kpc based on the apparent peaks in Fig.~\ref{VcircByType} -- occur for both rotation- and dispersion-dominated galaxies as defined above using the circularity measurements, at least for galaxies with stellar mass $>10^{9}\Msun$, although more careful work, especially with regard to the definition of a peak, will be required in the future.

From the right-hand columns, it is clear that at all masses, TNG galaxies are dark matter-dominated, i.e. $f_{\rm DM} (< r_{\rm half}) \gtrsim 0.5$ at the median $r_\rmn{half}$. The lowest fractions at the half stellar mass radius occur for galaxies in $10^{12}\Msun$ haloes, where the the star formation efficiency is highest. The $10^{11}\Msun$ mass bin shows evidence for the effect of limited resolution; it also predicts average DMFs that are only mildly dependent on galactocentric distance, with average DMFs in the range 0.75-0.87 in comparison to DMFs as low as 10-20  per cent in the centres of the most massive haloes ($\sim 10^{13-14}\Msun$) and rising to about 90 per cent outwards of $\sim 50-100$ kpc; all haloes in this sample with $M_\rmn{200c}>10^{12}\Msun$ show monotonically increasing DMFs. 

 We conclude that TNG achieves a balance between the dark matter and baryonic distributions that corresponds to flat rotation curves, and does so across a wide range of host halo masses. At the same time it also generates a diverse set of matter distributions at fixed halo mass. We will explore whether the individual subcomponents have the right distribution, as parametrised by the DMF, in Section~\ref{dis}.

\subsubsection{The case of $10^{12}\Msun$ haloes}

We expand on the matter distributions and the balance across components by focusing on $z=0$ MW-mass haloes in Fig.~\ref{VcircByType}, in which we also compare the TNG results to those from a model that assumes different baryon physics, namely Illustris. From top to bottom, we give the circular velocity curves of the 20 TNG100 $10^{12}~\Msun$ haloes from Fig.~\ref{VcircM} (left-hand column, second panel from the top) for the total matter (DM+b, top), dark matter (middle) and stars (bottom); we neglect the gas component as Fig.~\ref{VcircByType} demonstrates it to be subdominant. For each TNG100 halo, we identify its counterpart in the DMO run (TNG100-DM, black curves) and in Illustris (red curves), using the matching algorithms detailed in Section~\ref{sec:matching}. For the Illustris analogue galaxies we also show in the middle panel their own DMO counterparts, from the Illustris-Dark run, in grey. 

At this halo mass the models return different matter distributions, as partially anticipated in Fig.~\ref{M200Func}. Generally, the TNG100 full-physics total profiles are more concentrated than the DMO profiles, because of the contribution of baryonic material in the centres of the haloes (blue vs. black curves in the top panel). Moreover, all 20 of the Illustris galaxies show a rising total circular velocity curve in the region [3-10]~kpc, whereas the TNG100 galaxies on average exhibit flat curves in the same radial regime (top panel). This is due to a greater density of stars in TNG galaxies within 5~kpc (bottom panel) -- the average TNG100 $r_\rmn{half}$ being half that of Illustris. Note that at fixed $10^{12}~\Msun$ halo mass, TNG galaxies have nonetheless lower galaxy stellar masses than Illustris galaxies (see e.g. Fig. 11 of \citealt{Pillepich17b}), but here in our matched halo sample the Illustris analogue galaxies reside in haloes with masses on average lower than $10^{12}\Msun$, because of the effects quantified in Fig~\ref{M200Func}. All types of galaxies are shown in Fig.~\ref{VcircByType}, but the stellar velocity curves of TNG galaxies in the bottom panel would be on average less peaked within a few kpc if we were to select only disc-dominated objects.  

The dark matter mass within the whole radial range is also higher in TNG100 haloes than in their DMO and Illustris counterparts (middle panel): namely, the TNG galaxy physics model produces a contraction -- the drawing in from larger radii to smaller radii -- of the dark matter in the presence of baryonic physics and material. We have calculated the dark matter profile that would be expected for each of our TNG100 sample galaxies given adiabatic contraction \citep{Blumenthal86,Gnedin04} and plot the median as the dashed light-blue curve. This model does a good job at predicting the hydrodynamical result between 5 and 10~kpc, and is an underprediction of less than 10~per~cent at larger radii. Adiabatic contraction is therefore a reasonable cause of the change in the TNG100 dark matter profile. 

The dark matter contraction produced in the Illustris model is somewhat different. First, it should be noted that the TNG and Illustris dark-matter only sets (black vs. grey) are different for two reasons: the TNG100(-DM) and Illustris(-Dark) runs are performed with different cosmological parameters; moreover, the effective median mass of the analogue DMO samples are slightly shifted according to the findings of Fig.~\ref{M200Func}. By comparing the Illustris haloes (red curves) to their Illustris-Dark analogues (grey curves), it can be seen that Illustris dark matter haloes are slightly expanded within $\sim 8$~kpc -- i.e., dark matter has been moved to larger radii -- but contracted beyond the same radius in comparison to their DMO counterparts, although to a lesser degree than in TNG. The different behaviour of this model is replicated in the Illustris adiabatic contraction model (dashed pink curve), which consistently overpredicts the measured Illustris dark matter circular velocity profile. At face value, this result is at odds with the findings by \citet{Dutton16}, according to whom haloes in which star formation is more efficient exhibit higher levels of contraction rather expansion. In fact, this statement does not hold across different models, as Illustris galaxies are more efficient than TNG100's at producing stars and yet exhibit a smaller degree of contraction within the corresponding stellar half-mass radii. At radii larger than the baryonic galaxy, the agreement with \citet{Dutton16} improves (see Section.~\ref{sec:DMenhancement}). It is therefore apparent that careful future work will be required to tease out the physical processes that set the dark matter profile, and how these lead to differences between simulations across all relevant galactocentric distances.

\begin{figure}       
        \includegraphics[scale=0.43]{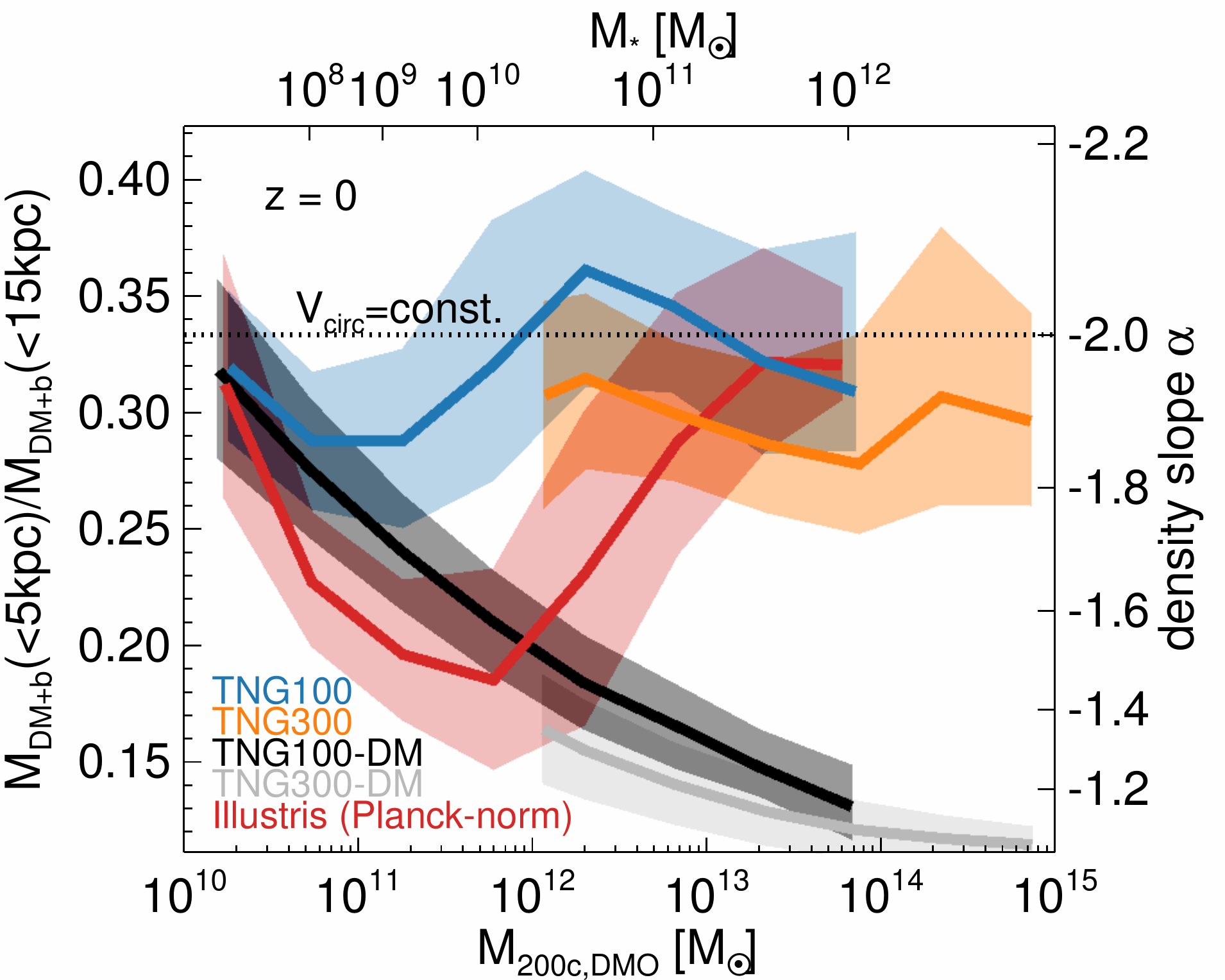}
 
\caption{$z=0$ enclosed total mass ratios for TNG100 (blue), TNG300 (orange) and Illustris (red), evaluated within the fixed apertures 5 and 15~kpc, as a function of the total halo mass from the DMO analogue measurements. A ratio of $\sim 0.34$ denotes flat circular velocity curves; a smaller (larger) value denotes rising (falling) curves. Thick lines denote median relations, and shaded regions encompass 68~per~cent of the galaxies at fixed halo mass. Here, the Illustris data have been renormalized by the ratio of TNG100-DM to Illustris-1-DM run results in order to take into account differences due to the change in cosmology. The black data in the top panel are from TNG100-DM, and the grey data are from TNG300-DM. The dotted horizontal line shows the value of the ratio that corresponds to a flat circular velocity curve. We include on the top $x$-axis of each panel values for the galaxy stellar mass corresponding to halo mass, based on the median $M_{200c}-M_{*}$ relation in TNG100, as calculated using galaxies within 10~per~cent of the stated values of $M_{*}$.}
\label{C5k15}
\end{figure}

\subsubsection{The flatness of TNG circular profiles}

Having examined the specific case of a subsample of $10^{12}\Msun$ haloes and qualitatively seen that TNG intermediate mass haloes have flatter or falling rather than rising circular velocity curves around their 3D stellar half mass radius, we extend the analysis to our entire selection of galaxies in the $M_{200c}=[10^{10},10^{14}]~\Msun$ range, where the halo mass is that of the DMO counterpart. We compute the change in the mass profile between two radii (Fig.~\ref{C5k15}), in order to quantify to what level different models predict flat circular velocity curves across our mass range.

We consider the ratio in enclosed mass between two fixed apertures: 5 and 15~kpc. The fixed aperture approach has the advantage that it is independent of any definition of galaxy sizes, but the disadvantage that it identifies different galaxy regions at different masses. Nevertheless, this aperture range is relevant for many different observations: for MW-mass objects, the $5-15$~kpc apertures bracket the position of the Sun, and for all $10^{12-13}\Msun$ haloes it encompasses the average 3D stellar half-mass radius. For more massive objects, the $5-15$~kpc galactocentric range is well inside the average 3D stellar half mass radius, where baryonic effects on the underlying dark matter -- if any -- are expected to be more relevant; for less massive objects, the range probes the outskirts of galaxies.

The aperture mass ratio results are shown in Fig.~\ref{C5k15} for TNG100 (blue), TNG300 (orange), Illustris (red), TNG100-DM (black), and TNG300-DM (grey) haloes, for the total (DM+b) mass. The mass enclosed ratios would equate to the value $\sim 0.34$ (horizontal dotted line) if the circular velocity curves (and, for disc galaxies, rotation curves) were perfectly flat across the probed $5-15$ kpc radial range. Hence, enclosed mass ratios lower (higher) than the dotted curve denote rising (falling) circular velocity profiles with galactocentric distance. On the right-hand $y$-axis, we also indicate what mass ratios are to be expected assuming a total mass density profile with average power-law slope $\alpha$ across the probed radial range: a flat circular velocity curve corresponds to an isothermal profile of slope $\alpha = -2$. Thick curves denote medians across all galaxies in $\pm 0.47$~dex halo mass bins, shaded areas encompass 68~per~cent of galaxies at fixed halo mass. To facilitate comparison, the Illustris data are multiplied by the ratio of the TNG100-DM and Illustris-Dark aperture ratios in order to remove the differences due to the change in cosmology.

As was hinted at in Fig.~\ref{VcircM}, within the TNG model, MW-mass haloes exhibit the flattest circular velocity curves in the $5-15$ kpc radial range. In fact, all but the least massive TNG100 galaxies are more concentrated than the DMO expectation (blue vs. black curves). For MW mass haloes, the median ratio of DM+b mass enclosed within 5 and 15~kpc is 50~per~cent higher than in the DMO simulation, and over twice as high for the very largest haloes. This relative increase in $<5$~kpc mass is sufficiently high that the ratio that corresponds to a flat circular velocity curve resides comfortably within the 68~per~cent of data region for $M_{200c}>10^{11}\Msun$. Within the same mass and radial range, the dark matter (only) circular velocity curves in DMO haloes would instead steeply rise inside-out, and more rapidly so the higher the mass. The TNG300 simulation gives qualitatively similar results, but with a decrement of $\sim0.03$ in aperture ratio due to the lower stellar masses \citep{Pillepich17b}, and also due to poorer resolution within 5~kpc as is apparent from comparing the two DMO simulations (black vs. grey). These results are in stark contrast to Illustris, which for $M_{200c}<10^{12}\Msun$ predicts galaxies on average \emph{less} concentrated (in total mass) than DMO haloes. Part of this shift can be explained by the fact that Illustris galaxies in that mass range have larger 3D stellar half mass radii than TNG galaxies at fixed mass.

\begin{figure}       
        \includegraphics[scale=0.68]{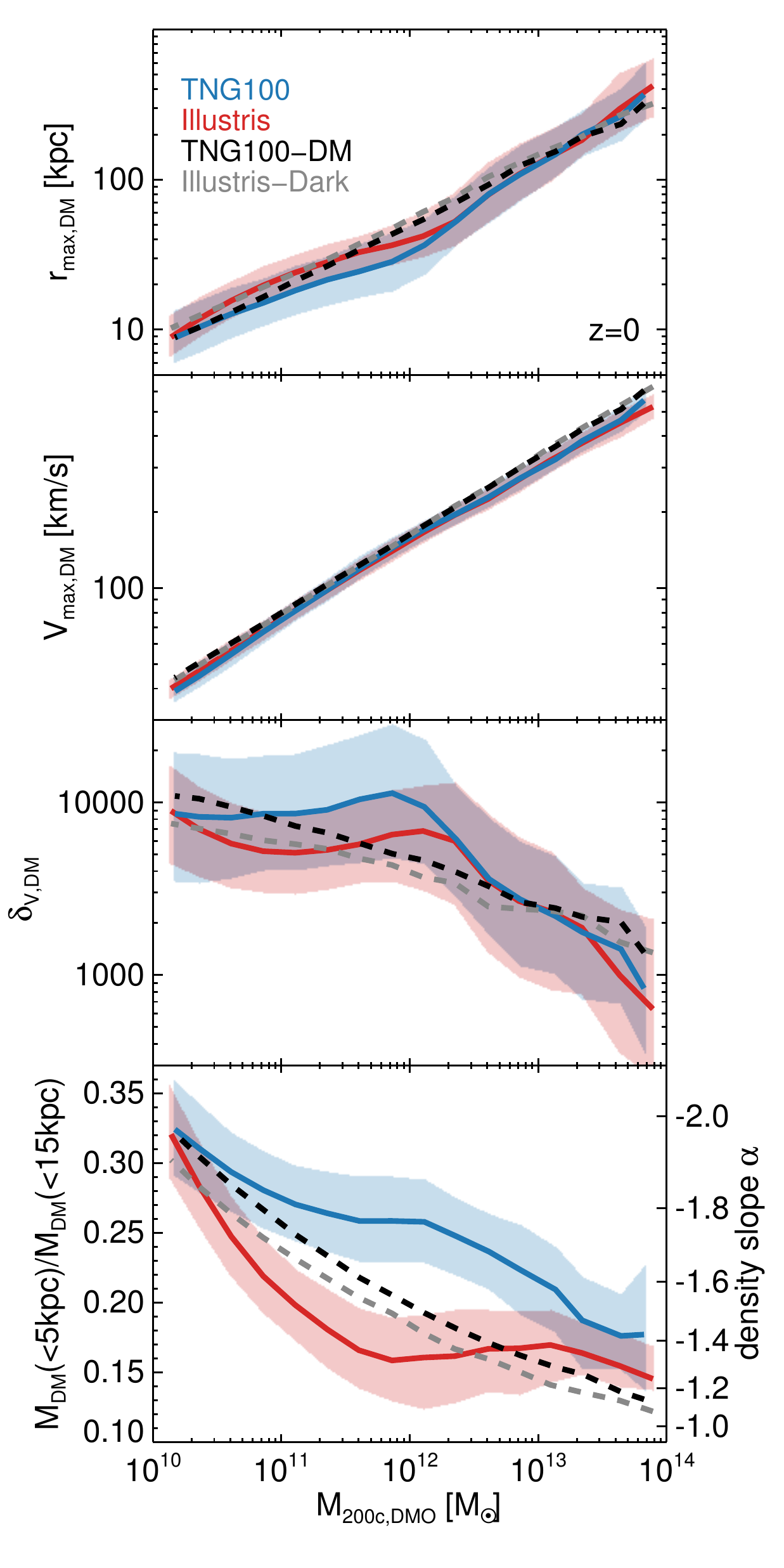}
        \caption{Dark matter halo properties as a function of halo mass (of the DMO counterparts), from top to bottom: $r_\rmn{max,DM}$, $V_\rmn{max,DM}$, the halo concentration $\delta_{V,\rmn{DM}}$, and the dark matter mass enclosed ratio between 5 and 15 kpc. Each of these quantities is calculated using the dark matter component alone. TNG100 is shown in blue and Illustris in red. Shaded regions mark 68~per~cent of the data and median relations are denoted with solid lines. The equivalent medians calculated with the DMO counterpart simulations are shown in black (TNG100-DM) and grey (Illustris-Dark).}
        \label{VmaxRmax}
\end{figure}

\subsubsection{Baryonic effects and dark matter enhancement}
\label{sec:DMenhancement}

The profiles seen thus far motivate us to quantify the degree to which baryonic physics and the presence of baryonic material alter the distribution of the underlying dark matter. We therefore focus on the {\it dark matter component alone} in Fig.~\ref{VmaxRmax}.

From top to bottom, we plot the average $r_\rmn{max,DM}$ and $V_\rmn{max,DM}$ of the DM circular velocity profiles, the dark matter halo concentration $\delta_{V,\rmn{DM}}$ (see Section \ref{meth}), and finally the ratio of the dark matter mass enclosed within 5 and 15 kpc, which is a proxy for the flatness of the dark matter circular velocity profile. All quantities are shown as a function of halo mass (of the DMO counterparts) and given as medians (thick curves) and 1-sigma halo-to-halo variations (shaded areas) for TNG (blue), its DMO counterpart TNG100-DM (black), Illustris (red), and its DMO counterpart Illustris-Dark (grey).

There is evidence for contraction in both the TNG and Illustris models: the TNG dark matter haloes are similarly or more concentrated than their DMO counterparts (blue vs. black) and more concentrated than their Illustris analogues (blue vs. red). While the peak values of the DM circular curves are within 10-15~per~cent from the DMO predictions (second panel from the top), TNG haloes have maximum radii ($r_\rmn{max,DM}$) up to 30~per~cent smaller than their DMO analogues (top panel), making overall their DM haloes more contracted than what is predicted by DMO simulations. The maximal change from TNG100-DMO for TNG100 is at $\sim 10^{12} \Msun$ and there is a similar, smaller suppression of $r_\rmn{max,DM}$ in Illustris at slightly higher masses. Both models show an increase in $r_\rmn{max,DM}$ of up to a few 10s of per~cent relative to TNG100-DMO for massive objects (a few $10^{13}\Msun$). 

The dark matter halo contraction is further assessed in the third panel from the top, where the simple dark matter halo concentration parameter $\delta_{V,\rmn{DM}}$ is up to a factor of 2 larger in TNG in comparison to the DMO haloes. The curves for Illustris and Illustris-Dark in this panel are similar to that reported in Fig.~6 of \citealt{Chua16}, who also used Illustris data but with a different definition of dark matter halo concentration, and confirm their findings.

Finally, the bottom panel demonstrates that the dark matter-alone circular velocity curves are flatter between 5 and 15 kpc in TNG haloes than in their DMO analogues. These ratios are consistent with dark matter density profiles with average power-law slopes in the 5-15 kpc apertures of about $-1.8$ for $10^{12}\Msun$ TNG100 (and TNG300) haloes, compared to about $-1.5$ for the TNG100-DMO analogue and the still shallower $-1.3$ for Illustris haloes. This steepening of the dark matter density profile slopes due to baryonic physics is consistent with the findings of \citet{Schaller16} from the APOSTLE simulations, although measured at somewhat different galactocentric distances. \citet{Dutton16} similarly found contraction of the haloes at 1~per~cent of the virial radius for haloes with high star formation efficiencies, that disappears or even transitions to expansion for low star formation efficiencies. Within our models, it is therefore plausible to expect that the dark matter density profile of our Galaxy should likewise be steeper than what is typically assumed for Navarro-Frenk-White haloes, and that the dark matter annihilation signal predictions \citep[e.g.][]{Springel08a} may be different than the DMO simulation-derived predictions across the studied mass range. 

We have verified with a comparison between the TNG100 and TNG300 curves (not shown) that the properties of the dark matter profiles are better converged than the total or baryonic matter distributions, despite the influence of baryonic physics on the dark matter.

\subsection{Dark matter fractions}

\subsubsection{$z=0$ averages within fixed apertures}

We now turn to the relative content of DM and baryonic mass and quantify the dark matter fractions (DMFs) as observables on their own. It is apparent from the results shown in the previous Sections that the effective DMFs within galaxies not only depend on how the baryonic mass is distributed, but also on how baryonic physics may alter the distribution of dark matter itself in comparison to DMO expectations.

\begin{figure}       
        \includegraphics[scale=0.57]{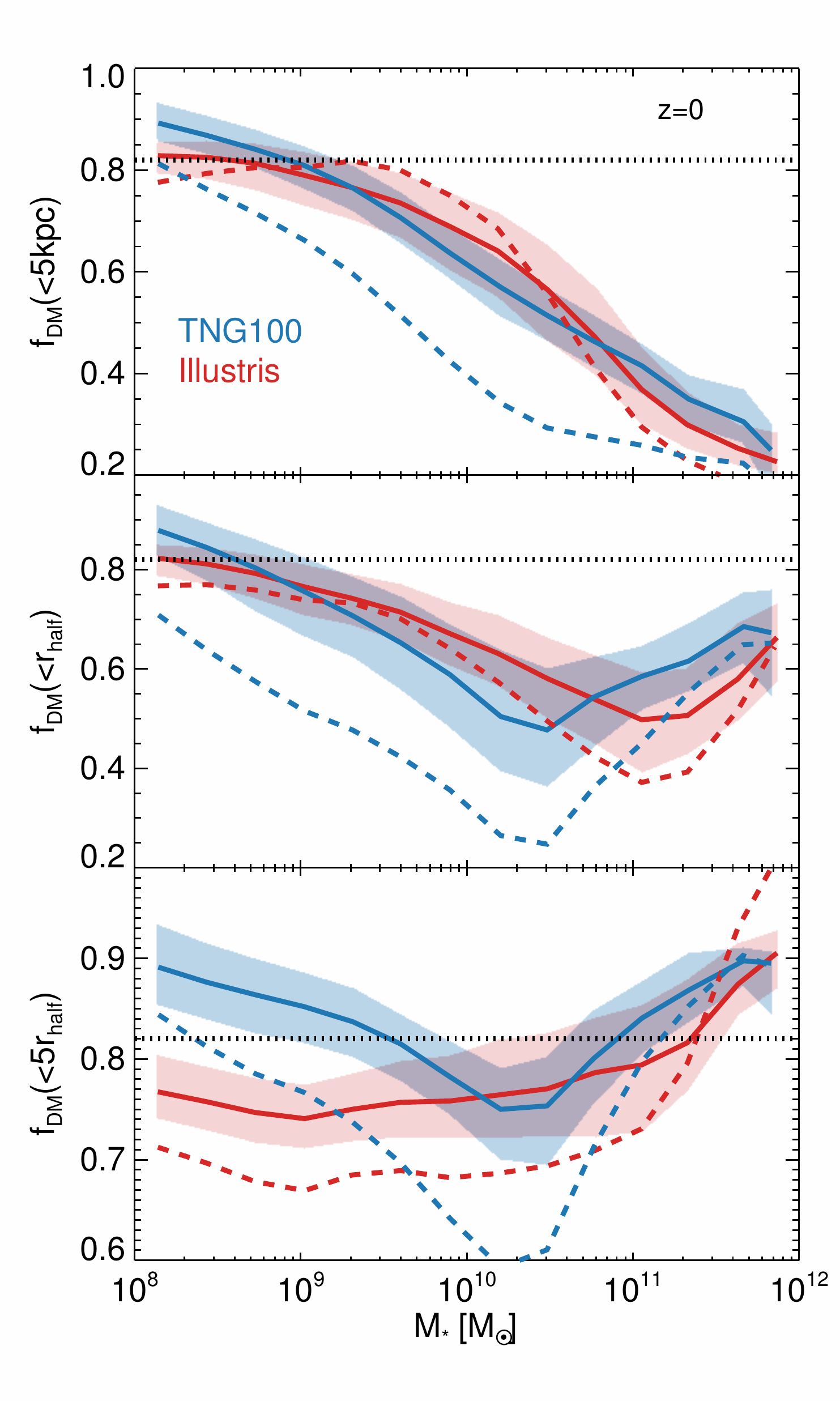}

        \caption{The mass fraction of dark matter (DMF) in TNG $z=0$ galaxies within a series of radii as a function of galaxy stellar mass: 5~kpc (top), $r_\rmn{half}$ (middle), and $5\times r_\rmn{half}$ (bottom).  The TNG100 and Illustris data are shown in blue and red respectively. As in previous Figures, the solid lines denote median relations and the shaded regions 68~per~cent of the data. The dashed curves denote the median results which would have been obtained if the galaxies simulated within the full-physics simulations (e.g. TNG100) were to be placed in their corresponding dark matter haloes simulated within DMO simulations (e.g. TNG100-DM), i.e. by neglecting the effects of baryonic physics on the distribution of dark matter.}
        \label{fDM}
\end{figure}

For each galaxy in our simulated samples, we measure the DMF within three characteristic radii -- 5~kpc, $r_\rmn{half}$ and $5r_\rmn{half}$ -- and we show the results in Fig.~\ref{fDM} for our TNG100 galaxies at $z=0$ as a function of galaxy stellar mass: blue thick curves and shaded areas for the running medians and 1-sigma contours, respectively. We also include the corresponding Illustris data with red curves and shaded areas.
The trend of DMF with galaxy mass depends on the chosen aperture: within fixed 5~kpc, more massive galaxies have lower DMFs ($0.25 \pm 0.05$ for $10^{12}\Msun$) than lower mass galaxies ($0.8\pm 0.05$ at $10^{9}\Msun$). Instead, within a mass-dependent half-mass radius aperture, the trend is {\it not} monotonic with galaxy mass. The TNG100 DMF within both $r_\rmn{half}$ and $5r_\rmn{half}$ is minimised at $3\times10^{10}\Msun$, reading respectively $0.5\pm0.1$ and $0.75\pm0.05$. This seems in good agreement with the idea that a few $10^{10}\Msun$ is the stellar mass at which galaxy formation is most efficient and, by extension, most strongly alters the dark matter profiles (as seen in Section \ref{sec:DMenhancement}). In fact, the rising of the DMFs at larger galaxy masses can also be a consequence of the shape of the galaxy size-mass relation, with the half-mass radii probing progressively farther galactocentric distances and hence farther out in the haloes as the galaxy mass increases. At all masses, the DMFs increase towards larger galactocentric distances (see again also right column in Fig. \ref{VcircM}).

\begin{figure*}       
        \includegraphics[scale=0.57]{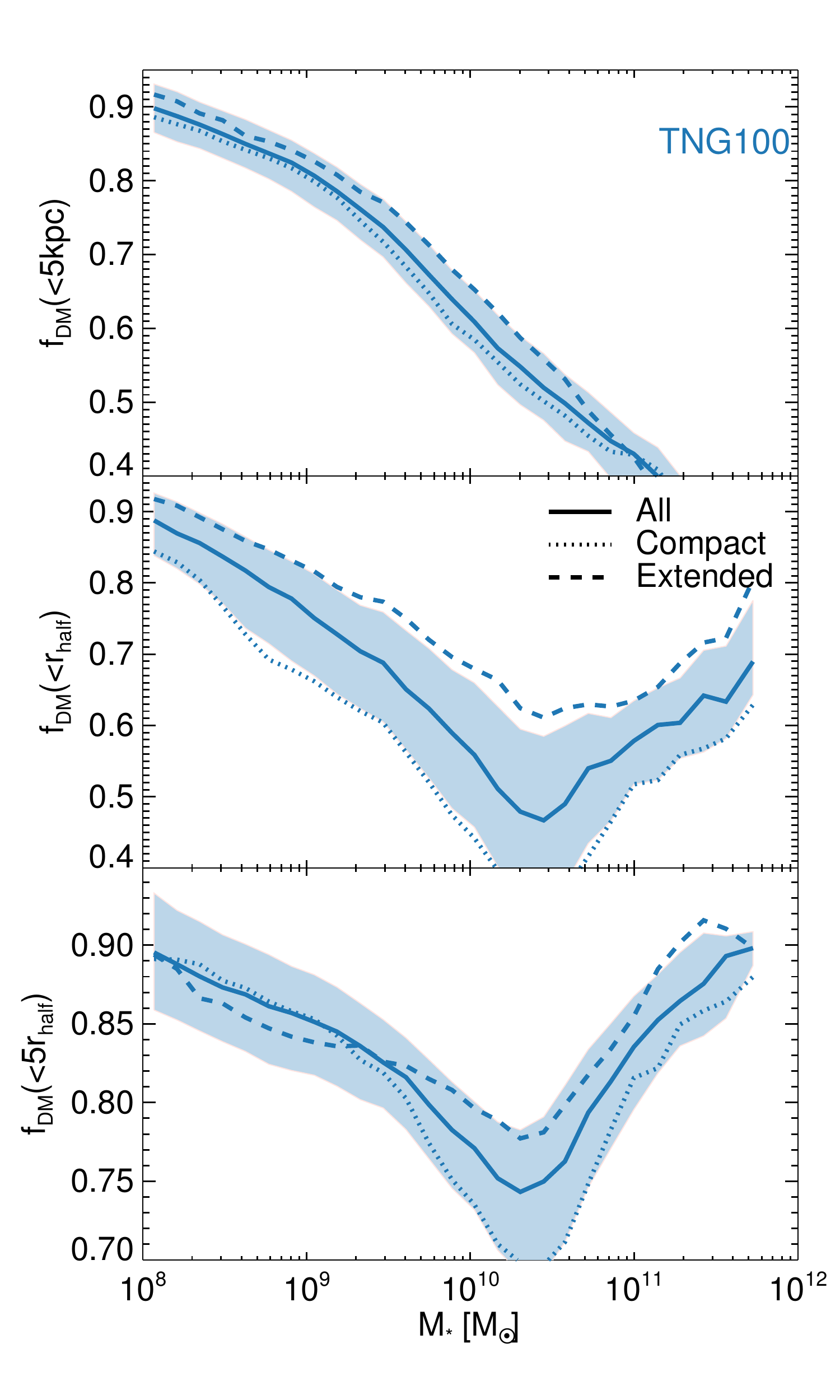}   		  
        \includegraphics[scale=0.57]{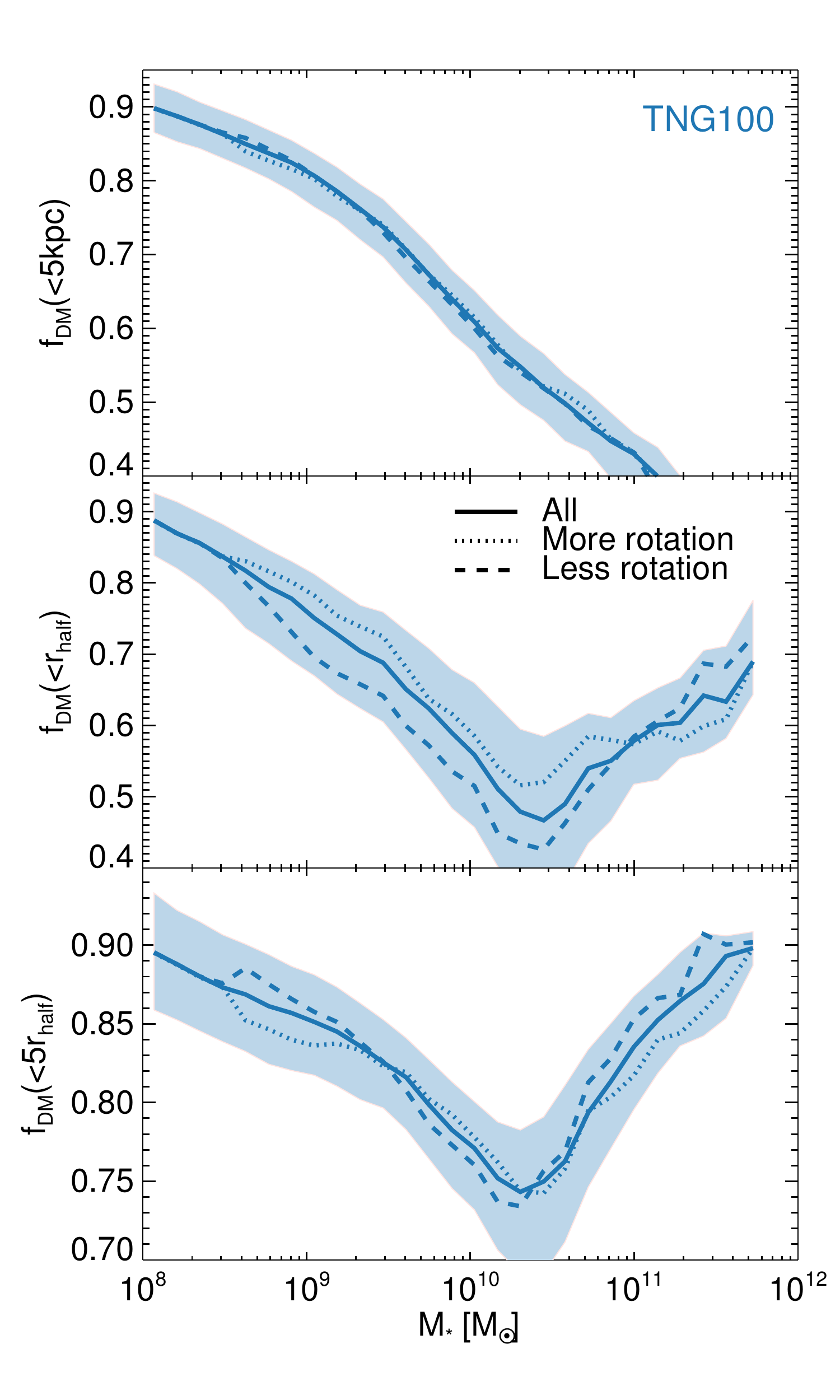}
        \caption{The median $z=0$ DMF within the same apertures as in Fig.~\ref{fDM}, as a function of stellar mass, separated by galaxy sizes (left) and kinematics/morphologies (right). The blue shaded regions denote 68~per~cent of the data. The median of the entire population of galaxies is shown as a solid line. In the left hand column the dashed and dotted lines show the medians of two subsamples of the data as split by size (25~per~cent highest $r_\rmn{half}$, dashed, and 25~per~cent lowest $r_\rmn{half}$, dotted, respectively). The solid line and the 68~per~cent region are replicated in the right-hand column, where the separation by subsamples is instead made by kinematics (25~per~cent highest disc fraction as dotted, and 25~per~cent lowest disc fraction as dashed.)}
        \label{fDM2}
\end{figure*}
In order to see how these results depend on the dark matter enhancement quantified in the previous Sections, we draw with dashed curves the DMFs we would obtain by painting TNG100 galaxies (with their baryon distributions) on their analogue DMO haloes from the TNG100-DM simulation: in practice, for each galaxy in the full-physics simulation we substitute the measured dark matter enclosed mass profiles with those of the corresponding DMO haloes. The measured TNG100 DMFs are up to twice the value of the accompanying DMO estimates, particularly around the peak of star formation efficiency at $\sim2\times10^{12}\Msun$. Otherwise, the TNG DMFs would be up to two times lower if baryonic physics did not induce a dark matter enhancement in the central regions of galaxies as is the case in TNG.

We have also examined curves from TNG300 and from the lower resolution versions of TNG100 (not shown), and they demonstrate that, at fixed galaxy mass, the TNG100 results of Figure~\ref{fDM} are converged to better than 10-15~per~cent, with the fully-converged TNG model predictions possibly being only slightly lower ($\la 10$~per~cent) than what the blue curves indicate.

Interestingly, within the inner regions of galaxies (upper panels), Illustris and TNG100 predictions for the DMFs are not so dissimilar, despite (or rather thanks to) some persistent but slight shifts associated to the differences in the stellar mass-halo mass relation between the two models. The difference between the {\it DMO-convolved} estimates of Illustris and TNG is large (dashed blue vs. dashed red): for $M_{*}<2\times10^{10}\Msun$ there is a much better agreement between the Illustris measurements and those obtained by painting Illustris galaxies into their DMO analogue haloes (red solid vs. red dashed) than for the TNG galaxies. Effectively, within the 5~kpc apertures, the presence of the galaxy in the Illustris model has smaller effects on the dark matter content of the inner halo than in the TNG model. An excess in dark matter in comparison to DMO predictions slowly emerges in Illustris towards higher stellar masses and larger distances (bottom panel).
 
Finally we note that the dark matter fractions in Illustris had already been quantified in \citet{Xu17} for a selection of massive ($M_{\rm stars} \gtrsim 10^{11} \Msun$) early type galaxies in the $z=0.1-1$ range. Their {\it projected} dark matter fractions within the effective radius read on average $0.4-0.6$, in broad agreement with our findings. We hence investigate in what follows the dependence of TNG DMFs on galaxy properties and redshifts.

\subsubsection{Dependence on galaxy and halo properties}

What produces the galaxy-to-galaxy variations in DMF at fixed stellar mass (Fig.~\ref{fDM}, shaded areas) that we see in TNG objects? By focusing on TNG100 only at $z=0$, we split galaxies by stellar sizes and morphologies and show results in Fig.~\ref{fDM2}. 

In the left-hand column, we bin our galaxies in stellar mass, split the sample in each bin into quartiles of the 3D stellar half mass radius distribution ($r_\rmn{half}$) and calculate the median DMF of the most compact 25~per~cent and most extended 25~per~cent galaxies, i.e. the upper and lower quartiles of the $r_\rmn{half}$ distribution. The stellar size of the galaxy has an apparent measurable effect on the DMF at fixed stellar mass (lower panels), with more extended galaxies exhibiting larger fractions of dark matter within $r_\rmn{half}$. This is to be expected when the DMF is measured within some multiple of the stellar half-mass radius, provided that the dark matter halo contraction does not correlate with galaxy size to compensate, e.g. smaller galaxies producing stronger halo contraction and bringing more dark matter inside $r_\rmn{half}$. The increase in DMF with effective radius has been found in dynamical analyses of local galaxies and $z\sim0.2$ lenses, and the correlation is even stronger with the enclosed dark matter density \citep{Tortora10,Tortora12}. Within $r_\rmn{half}$, the difference in DMF between galaxies in the lowest and highest quartiles in $r_\rmn{half}$ is largest around $3\times10^{10}\Msun$ -- the approximate peak of star formation efficiency -- with a shift of more than $\sim50$~per~cent. Towards higher and lower masses the difference shrinks, particularly at the faintest galaxies considered. A similar behaviour occurs at $M_{*}>3\times10^{10}\Msun$ in the $5r_\rmn{half}$ radius to a smaller degree, but disappears at lower radii. 

The DMFs measured within the fixed 5~kpc aperture show similar behaviour although to a smaller degree than in the lower panels, with higher DMFs for more extended galaxies. This is because some fraction of the fixed stellar mass is redistributed from within the aperture to outside for more extended galaxies, thus increasing the DMF within the aperture; note that if instead contraction of the halo for smaller sizes dominated over the redistribution effect, we would see the opposite effect where smaller galaxies had higher DMFs\footnote{It will be possible to check whether smaller galaxies produce more contracted haloes than larger galaxies of the same mass by comparing their dark matter distributions to those of the DMO simulation counterparts.}. The 5~kpc radius is larger than the typical galaxy size, at least up to $\sim 10^{10}\Msun$ in stars and it gets quickly much smaller than the typical sizes of more massive galaxies. It is in the intermediate regime, around $10^{10}\Msun$, where $r_\rmn{half}\sim5$~kpc, and also where the scatter due to $r_\rmn{half}$ is strongest. The small overall systematic effects in the top panel, at essentially all masses, suggest that the larger effects in the lower panels are mostly driven by the change in effective aperture where the dark matter fraction is probed for more compact or more extended galaxies -- larger sizes at fixed stellar mass probe further out into the halo, hence larger DMFs -- rather than a strong change of the matter profile ratios.

The effect of the galaxy sizes is expected to be translated into the relationship between DMF and morphology, given that TNG elliptical/quiescent galaxies are more compact than disc/star-forming galaxies of the same stellar mass \mbox{\citep{Genel17}}. Here we approximate morphology with the internal kinematics, and comment later on how the two differ; we show the results for a kinematic decomposition of more rotation/less rotation in the right hand panel of Fig.~\ref{fDM2}. The median DMF within $r_\rmn{half}$ as a function of $M_{*}$ for less- (dashed curves) and more strongly (dotted curves) rotating galaxies (mid right panel) mirrors the gap between the lower quartile and upper quartile radius galaxies, although to a lower degree, particularly around the star formation efficiency peak and somewhat at lower stellar masses. At larger galactocentric distances (bottom right panel) these features are not clearly replicated, nor for fixed physical apertures\footnote{However, the splitting between galaxy populations in the left and right hand side of Fig.~\ref{fDM2} is by construction different and selects different subsets of the galaxy distribution, hence a quantitative direct comparison of the effects of sizes vs. morphologies is not viable.  Moreover, low-rotation high mass galaxies are likely to be ellipticals whereas low-rotation low mass galaxies tend instead to be irregular in morphology.}.

The trends described in Fig.~\ref{fDM2} do account for much of the $\pm 1$~$\sigma$ scatter, which here is reported as the blue shaded area. 
We have also studied other halo and galaxy properties to see whether they may be responsible for larger or smaller DMFs at fixed stellar mass than the average, either at the intrinsic level or due to their role in setting the galaxy size. We find that the virial halo mass, $M_{200c}$, correlates slightly with DMF at the extreme high stellar mass end, with minimal impact around the star formation peak, and is {\it inversely} correlated with DMF for lower mass galaxies, including within the 5~kpc fixed aperture. The halo concentration (see Section~\ref{meth}) has a negligible effect on the fixed aperture but instead induces $\sim$20~per~cent scatter in the $1\times r_\rmn{half}$ aperture at the star formation peak, $10^{10}\Msun$, which suggests it is connected to the setting of the galaxy sizes. Finally, we find that the strongest source of scatter in {\it low-mass} galaxies, especially within the fixed aperture, is the star formation rate: higher star formation rates correlate with lower DMFs, but this relation is reversed for galaxies in $10^{12}\Msun$ haloes. There is therefore a rich interplay between the galaxy and its host that sets the DMF, and we do not examine further whether any of these subpopulations exhibit higher or lower degrees of contraction; a future study will examine the physical connection of these processes in contrast to the empirical approach adopted here.

\begin{figure}
 \includegraphics[scale=0.55]{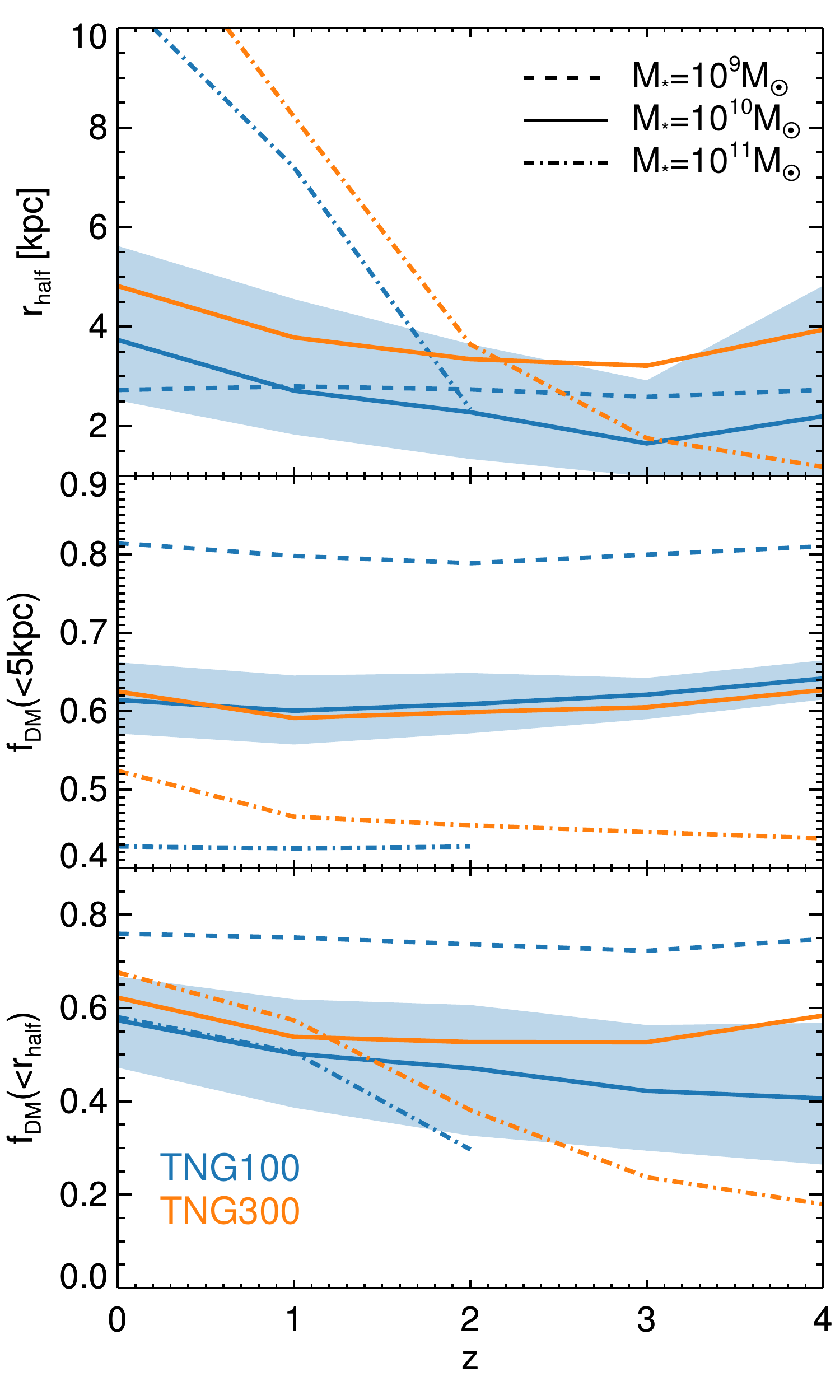}%
 \caption{Galaxy properties in TNG100 and TNG300 as a function of redshift. Top panel: the evolution of $r_\rmn{half}$, middle panel: dark matter fraction within 5~kpc (physical), bottom panel: dark matter fraction within $r_\rmn{half}$.  Results are shown for three stellar mass bins: $10^{9}$ (dashed line, TNG100 only), $10^{10}$ (solid line) and $10^{11}\Msun$ (dot-dashed line). Each stellar mass bin contains galaxies within 10~per~cent of the stated stellar mass.} 
    	\label{RHvZPlus}
\end{figure}
\subsubsection{Redshift evolution}

The properties of $z=0$ galaxies are influenced by the state of galaxies at earlier redshift. We therefore measure the DMFs within $r_\rmn{half}$ and 5~kpc (in physical kpc) at redshifts 0, 1, 2, 3 and 4, and show the results along with the evolution of $r_\rmn{half}$ itself in Fig.~\ref{RHvZPlus}: we focus on TNG100 and TNG300 galaxies and on three stellar mass bins: $10^{9}$, $10^{10}$ and $10^{11}\Msun$ with a bin width of $\log_{10}(r_\rmn{half})=0.083$. Note that we do not include a $10^{9}\Msun$ bin for TNG300: this is due to poor spatial resolution of these objects in this simulation. 

The redshift evolution of the DMFs in galaxies depends again on the choice for the aperture over which they are measured. The DMFs within the fixed aperture of 5~kpc change by no more than 10~per~cent between $z\sim4$ and today, for all our models and stellar masses. A weak trend can be seen for $10^{10}\Msun$ galaxies, whose $f_\rmn{DM}(< 5~ {\rm kpc})$ slightly decrease towards low redshifts. 

However, galaxy sizes at fixed stellar mass increase as redshift decreases, with stronger trends the larger the galaxy mass. Consequently, the DMFs measured within $r_\rmn{half}$ increase towards lower redshifts at fixed galaxy mass, or expressed differently DMFs are smaller at higher redshifts than today because the baryonic material is more centrally concentrated at higher redshifts than today. This is particularly true for the $10^{11}\Msun$ mass bin galaxies, which are so concentrated at high redshift that their $r_\rmn{half}$ -- although likely not their total extent -- is smaller than that of lower mass galaxies; this evolution has been shown in a series of previous simulations \citep{Hilz12,Hilz13,Remus17} as well as more recently by \citet{Kuiler18}, and is well attested from $z\sim2$ in observational studies (\mbox{\citealp{ForsterSchreiber09,Toft12}}; \mbox{\citealp{Tortora14,Genzel17}}), and we will expand on the observational comparison in Section~\ref{sec:Ddgat}. Finally, we note that the TNG300 galaxies are consistently more extended than their TNG100 counterparts, and thus have higher DMF within this aperture. This is because galaxy sizes are not fully converged at TNG300 resolution, are larger than in TNG100, and these resolution trends have been explored in \citet{Pillepich17} (Fig.~A2 and related discussion). By contrast, the DMF measured within 5~kpc for the $10^{10}\Msun$ bin shows little difference between the two simulations.

\subsection{Model variations}
\label{sec:ModVar}
\begin{figure}
\includegraphics[scale=0.33]{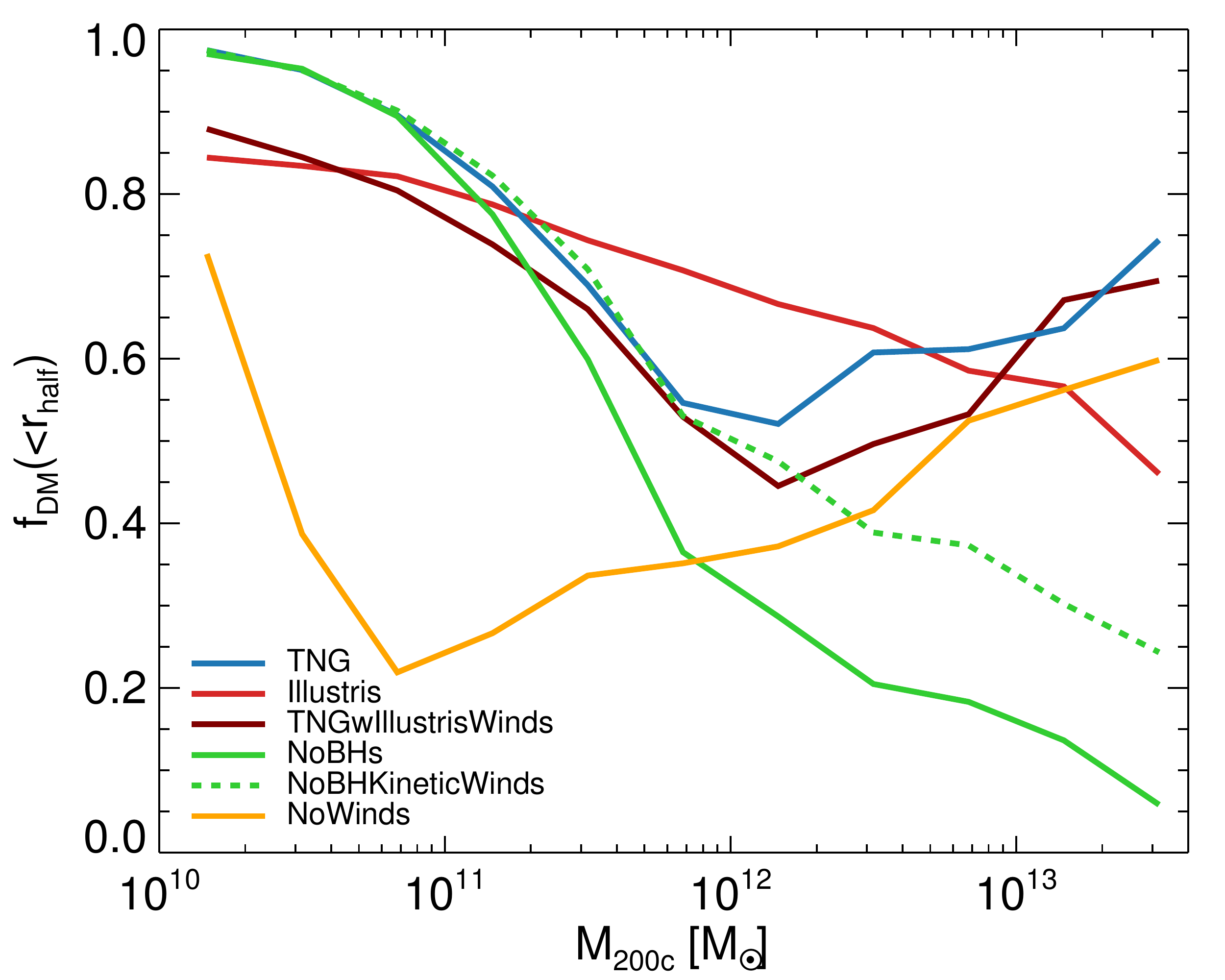}

 \caption{ DMF within $r_\rmn{half}$ as a function of $M_{200c}$ at $z=0$ for a range of model variations. The six models are TNG (blue), Illustris (red), TNG with Illustris winds (maroon), no black holes (green, solid), no black hole kinetic winds (green, dashed) and no winds (orange), all performed in the same $25 h^{-1}$ Mpc volume, which is a much smaller box than the flagship simulations used thus far.}

    	\label{MODELS}
\end{figure}

In the previous Sections we have often contrasted our principal results emerging from the TNG100 and TNG300 simulations with other different or simplified numerical experiments, particularly the Illustris simulation and the gravity-only versions of both TNG and Illustris galaxies.  While the agreement in the DMF-stellar mass relation between TNG100 and Illustris may be interpreted as a robust confirmation of the prediction, such inter-model consistency is instead somewhat coincidental. We expand on this in Fig.~\ref{MODELS}, where we show the median DMFs of TNG and Illustris galaxies now as a function of halo mass and in comparison to a series of other TNG model variations. There, we switch off or modify one at the time individual fundamental features of the TNG galaxy model: no galactic winds i.e. no stellar feedback; no black holes (BHs) and their feedback; no low-accretion rate feedback from BHs, which in TNG is implemented as a kinetic BH-driven wind; the TNG model where the galactic wind implementation reproduces the one adopted in Illustris.   
These models were performed at TNG100 resolution in the same $25~h^{-1}$Mpc volume \citep[see ][for more details]{Pillepich17}, therefore we plot in Fig.~\ref{MODELS} TNG and Illustris models also run in this smaller box. 

Interestingly, all model variations in Fig.~\ref{MODELS} agree that the DMF within galaxies residing in the least massive haloes included in our sample ($\sim 10^{10}\Msun$) is large, $0.8-0.9$: this is the case also for the ``NoWinds'' run, which returns galaxy stellar masses by many factors larger in comparison to the fiducial model and observations (see Fig.~8 of \mbox{\citealt{Pillepich17}}). Both of the BHs-related modifications (green) return much lower DMFs than our fiducial model at a few $10^{11}\Msun$ halo masses and above, consistent with the fact that they overestimate the galaxy stellar masses and have lower-than-required galaxy sizes (see again Fig.~8 of \citealt{Pillepich17}). In fact, the case that by far departs most strongly from the TNG fiducial model in terms of galaxy properties is the one where the TNG galactic winds implementation is substituted with the Illustris winds implementation (dark red): its values for the stellar mass-halo mass (SMHM) ratio  at a halo mass of $10^{12}\Msun$ are many factors larger than in TNG (see Fig.~9 of \citealt{Pillepich17}), and the galaxy sizes are twice as large as is the case in TNG. However, it returns a not too dissimilar DMF-halo mass relation compared to TNG. The non-monotonic trend of the latter would indeed appear to be strongly connected to how peaked and narrow the SMHM relation is and how rapidly the star formation efficiency drops for larger halo masses, modulo the shape of the galaxy size-mass relation as well. The minimum in DMFs occurs in some cases where the SMHM ratio is maximal: for the ``NoWinds'' case (yellow), for example, the latter is shifted to much lower halo masses than TNG. However, a break in monotonicity may not be present at all if the SMHM ratio is not peaked enough, as is the case for the ``NoBHs'' (green solid) and ``NoBHKineticWinds'' (green dashed) runs, and to a lesser degree for the Illustris one.

We have already shown in Fig.~\ref{VcircByType} that the contraction in $10^{12}\Msun$ TNG100 haloes is well described by adiabatic contraction models, and that the corresponding Illustris dark matter profiles are not.
We postpone to future analyses the task to quantify the level of dark matter contraction or expansion in these alternative implementations, and to identify the physical reasons responsible for the dark matter halo contraction that emerges in TNG galaxies. However, we conclude by reminding the reader that all the alternative implementations to TNG and Illustris models discussed in Fig.~\ref{MODELS} are securely ruled out by observational constraints on the $z=0$ galaxy stellar mass function and galaxy size-mass relation: this is the case for the ``TNGwIllustrisWinds'' run too, even though it gives DMFs somewhat consistent with TNG. For the Illustris simulation, the halo mass trend of DMFs within $r_\rmn{half}$ compares to TNG differently than the trend with galaxy mass, with the latter (former) being (not) monotonic with mass. With a lower degree of dark matter halo contraction at the relevant radial scales (Figs.~\ref{VmaxRmax}, \ref{fDM}), the Illustris model produces similar DMF predictions to TNG at fixed galaxy stellar mass. However, as outlined already, Illustris galaxies are more massive and more extended than TNG galaxies and in worse agreement with observational constraints than TNG \citep{Pillepich17b,Genel17}: this makes overall the TNG simulations a more plausible and reliable theoretical benchmark for comparison to observations of other galaxy and halo properties.

\section{Discussion and Implications}
\label{dis}
\begin{table*}
    \caption{Table of observations with measured DMFs. We include the paper in which the observations are published, the name of the parent survey if applicable, the redshift, $z$, of the sample, the type of galaxies included, a brief description of how the galaxies were selected, the IMF assumed/determined in each study, the type of aperture within which the DMFs were calculated, and the value/range of the inferred DMFs.  The \citet{IPB15} and \citet{BlandHawthorn16} papers collate data from other studies that had already derived an IMF: therefore, in these two cases the IMF is written in brackets.}
	\begin{tabular}{p{2.3cm}p{1.2cm}p{1.0cm}p{1.7cm}p{3.0cm}p{1.7cm}p{1.7cm}p{1.7cm}}
\hline
&&&&&&\\
        Paper & Survey & z & Galaxy types & Galaxy selection & IMF & DMF aperture  &  DMF\\                
&&&&&&&\\
\hline
&&&&&&&\\
\citet{IPB15} & N/A & 0.0 & Milky Way & Milky Way & (MAP) & [0,10]~kpc & $<0.6$ \\
         \citet{Wegg16} & N/A & 0.0 & Milky Way & Milky Way &  Kroupa &[0,10]~kpc & $<0.5$ \\
        \citet{BlandHawthorn16} & N/A & 0.0 & Milky Way & Milky Way & (Kroupa) & [4.5-12]~kpc & $0.45$ \\
         \citet{Bovy13} & N/A & 0.0 & Milky Way & Milky Way &  MAP & 2.2$R_\rmn{d}$ & $0.3$ \\
&&&&&&&\\
\hline
&&&&&&&\\ 
\citet{Alabi17} & SLUGGS & $<0.007$ & Early-type & $\sim L_{*}$, representative of local early-type population &  Kroupa & 5 projected half-light radii & [0.1, 0.95] \\
   
       \citet{Zhu14} & N/A & $\ll0.01$ & M87 & M87 &  Constant M/L & Matter density as function of radius & $<0.3$ (20~kpc) \\
      \citet{Cappellari13} & ATLAS$^\rmn{3D}$ & $<0.01$ & Early-type & $M_{*}>6\times10^{9}\Msun$, no spiral arm or dust lane & Variable & Projected half-light radius & $\le0.4$ \\   
      \citet{Wojtak13} & N/A & $<0.1$ & Early-type & ... &  Chabrier & 5 projected half-light radius & $\ge0.5$ \\
   \citet{Tortora12} & SPIDER & $<0.1$ & Early-type & ... & Chabrier & Projected half-light radius & $\le0.5$ \\ 
       \citet{Barnabe11} & SLACS & $<0.33$ & Early-type & ... & Chabrier/ Salpeter & De-projected half-light radius & $\le0.5$ \\ 
&&&&&&&\\
\hline
&&&&&&&\\ 
\citet{Courteau15} & DiskMass &$<0.01$ & Discs & $i\sim30^{\circ}$, $100<V_\rmn{rot}<250$~\kms & Kroupa & $2.2R_\rmn{d}$ & [0.5, 0.9]\\
        \citet{Courteau15} & SWELLS & [0.08,0.2] & Edge-on spirals & SLACS galaxies w/ lensing & Salpeter for bulge, Chabrier for disc & $2.2R_\rmn{d}$ & [0.1, 0.4]\\   

        \citet{Genzel17} & SINS/zC-SINF+ KMOS$^{3D}$ & [0.9,2.4] & Star-forming field galaxies & $M_{*}>10^{9.6}\Msun$, $\pm0.6$~dex around star-forming main sequence, disc & Chabrier & Best-fit half-light radius (optical light) & $\le0.22$ \\
         \citet{Kranz03} & N/A & $\ll0.01$ & High surface brightness, late type spirals & Spirals, $M_{*}=[1-10]\times10^{10\Msun}$ & Assumes colour-M/L relation & 2.2$R_\rmn{d}$ & $[0-0.5]$\\
        \citet{Fragkoudi17} & N/A & $\ll0.01$ & NGC1291 & NGC1291&  Fits M/L & 2.2$R_\rmn{d}$ & $0.25$\\
&&&&&&&\\ 

       \hline
	\end{tabular}
    	\label{ObsTab}
\end{table*}

We conclude our study by putting the TNG findings into the broader context: in particular we attempt a zeroth-order comparison to the  dark matter fractions within galaxies inferred from observations. In the following, we proceed by juxtaposing the theoretically-derived DMFs to the observationally-derived ones {\it at face value}, namely we do not retrace the steps of the observational analyses on virtual observations of our simulated galaxies. We postpone this task to future work. 

We compare theory vs. observation results for our own MW, higher-mass galaxies, different morphological types, and finally for higher-redshift galaxies. In Table~\ref{ObsTab}, we collate a series of observations related to these galaxy samples of different morphological types and redshifts. The inferred DMFs from these observations (in all cases, de-projected 3D mass fractions, even if the aperture is projected in 2D) are reported in the right-most column, denoting a large diversity across galaxies and apertures, but also possibly non-negligible, method-dependent systematic uncertainties for galaxies of similar types and mass. There we highlight the redshift of the sample, the type of galaxies included, a brief description of how the galaxies were selected, the IMF assumed or determined in each study, and the type of aperture within which the DMFs have been derived, i.e. whether 3D or 2D projected apertures.
We compare the results from the TNG100 simulation to representative subsamples of the observational data for the MW in Figs.~\ref{BOVY} and \ref{BOVY2} \citep{Bovy13,IPB15}; for early-type objects in Fig.~\ref{SLUGGS} \citep{Alabi17,Cappellari13, Barnabe11,Tortora12,Wojtak13,Zhu14}; the recent $z=2$ \citet{Genzel17} study and a few sets of $z=0$ late-type data \citep{Kranz03,Barnabe12, Dutton13, Martinsson13a,Martinsson13b,Fragkoudi17} in Fig.~\ref{GENZEL}.

\subsection{The Milky Way rotation curve}

The good agreement with observations of TNG galaxy integral properties (e.g. stellar masses, sizes, but also colours and others -- see Section \ref{sims}) lends credibility to the matter distributions within galaxies predicted by TNG and shown in Section \ref{res1}. There we have quantified to what degree, and at what masses and galactocentric distances, the falling (stellar) and rising (dark matter) components of the circular velocity curves balance each other to return flatter total circular velocity profiles (Fig.~\ref{C5k15}). Here, we therefore investigate how consistent the matter profiles of TNG MW-like galaxies are with estimates of Galaxy rotation curves.

The MW is a crucial and fundamental test bed for constraining the nature of dark matter, and numerical simulations have been routinely used to provide estimates for the local dark matter density and velocity dispersion needed in direct detection experiments \citep[e.g.][]{Ling10,Pillepich14, Bozorgnia16, Butsky16,Kelso16,Sloane16, Bozorgnia17}, for the inner slope of the dark matter density profile and hence for the expected signals of dark matter annihilation \mbox{\citep{Springel08a,Schaller16,Calore15}} or decay \citep{Lovell15} from the Galactic Centre. The sole comparison mentioned in this work between TNG and Illustris MW-like haloes (Fig.~\ref{VcircByType}) showcases the current level of systematic uncertainties that affect model predictions about CDM. However, some of these uncertainties can often be solved comfortably by validating or discarding the models holistically based on how they perform against an ever broader set of observational findings. Given that our simulations constitute the most recent, and thus hopefully, the best informed by observations and theory iteration of this endeavour, our results will have an important contribution towards the detection (or otherwise) of dark matter in our own MW.

\begin{figure*}
 \includegraphics[scale=0.84]{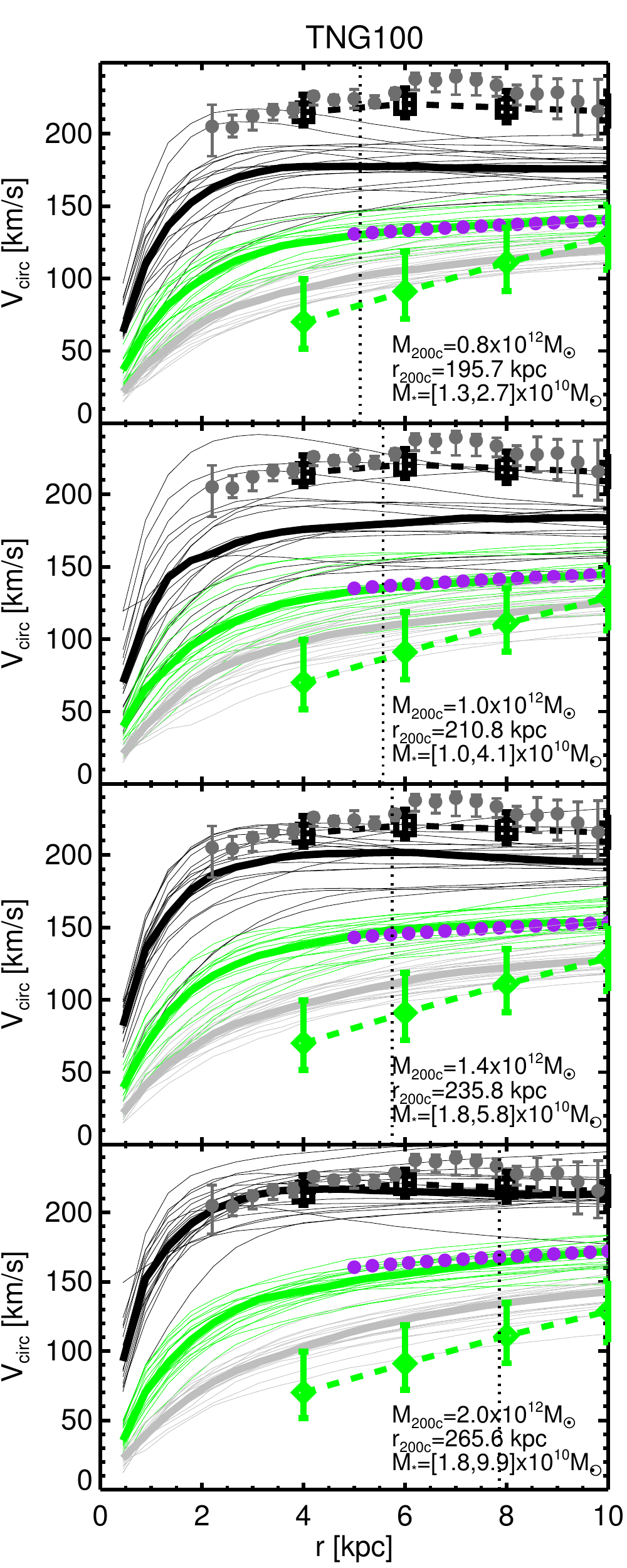}
  \includegraphics[scale=0.84]{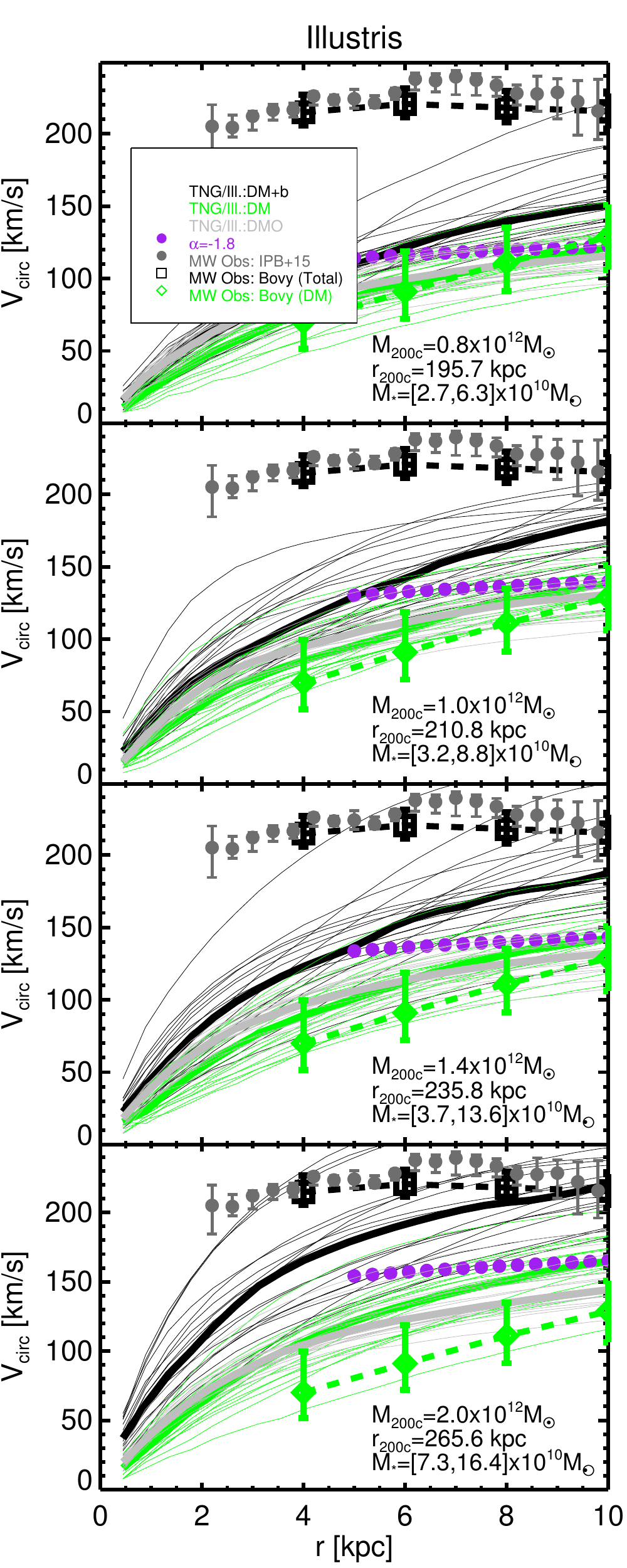}

\caption{Circular velocity profiles of $z=0$ MW-like galaxies from TNG100 (left) and Illustris (right) for halo masses 0.8, 1.0, 1.4 and $2.0 \times10^{12}\Msun$, from top to bottom. Each panel contains the 20 galaxies closest to the target mass. The range of stellar masses in each halo mass bin is given in each panel. We show the theoretical dark matter in green and the sum of dark matter and baryons in black. We include DMO counterparts as grey curves. As previously, thin curves denote individual galaxies and thick curves the median of the population. The circular velocity profile corresponding to a dark matter density profile with power-law slope equal to $\alpha=-1.8$ in the 5-15 kpc range is shown as purple dots for reference. Observationally derived constraints include: \citet{Bovy13} in open squares and diamonds for the total and dark matter alone profiles of our Galaxy, and the median rotation curve compiled by \citet{IPB15} as closed, grey circles. The \citet{Bovy13} error bars are derived as 68~per~cent confidence limits, and the \citet{IPB15} error bars denote 68~per~cent of the data points, and therefore do not take account of the errors on the individual points.} 
    	\label{BOVY}
\end{figure*}

The MW halo mass has been suggested to fall anywhere in the range of $5\times10^{11} - 2\times10^{12}\Msun$, depending on the choice of method and the definition of the MW halo mass itself \citep{Kahn59, Sales07a, Sales07b, Li08,Busha11, Deason12,Wang12, Gonzalez13, BK13, Cautun14b, Penarrubia14,Piffl14,WangW15}. Complicating the issue is the presence of the Large Magellanic Cloud (LMC), which may itself be embedded in a subhalo of mass $2\times10^{11}\Msun$ just 50~kpc from the MW centre \mbox{\citep{Penarrubia16}}.  For the purposes of this comparison, we retain $M_\rmn{200c}$ as our preferred definition of halo mass, and consider four possible halo masses that span the range described above: $M_{200c}$ --  $8\times10^{11}$, $1\times10^{12}$, $1.4\times10^{12}\Msun$, and $2\times10^{12}\Msun$ -- and select the 20 disc-dominated (circularity-defined disc fractions $>0.7$) simulated galaxies with masses closest to each of these four halo masses at $z=0$\footnote{Since we find that only a small number of our $8\times10^{11}\Msun$ curves make contact with the observational data, we elect to choose this as our lowest mass halo bin rather than $5\times10^{11}\Msun$}. In Fig.~\ref{BOVY}, we compare the simulated circular velocity curves of MW-like galaxies to two observationally-derived datasets: the Galaxy mass model of \citet{Bovy13}, which returns both the total and dark matter alone rotation/circular velocity profiles (square and diamond symbols), and the total rotation curve data compiled by \mbox{\citet{IPB15,Pato17}} from sources that include masers, gas kinematics and better stellar kinematics \citep[e.g.][]{McClureGriffiths07,Battinelli13,Reid14}, grey filled circles. Results are shown out to a radius of 10~kpc for TNG100 (left) and Illustris for comparison (right), including profiles from the DMO analogue haloes.

\begin{figure*}
 \includegraphics[width=15.0cm]{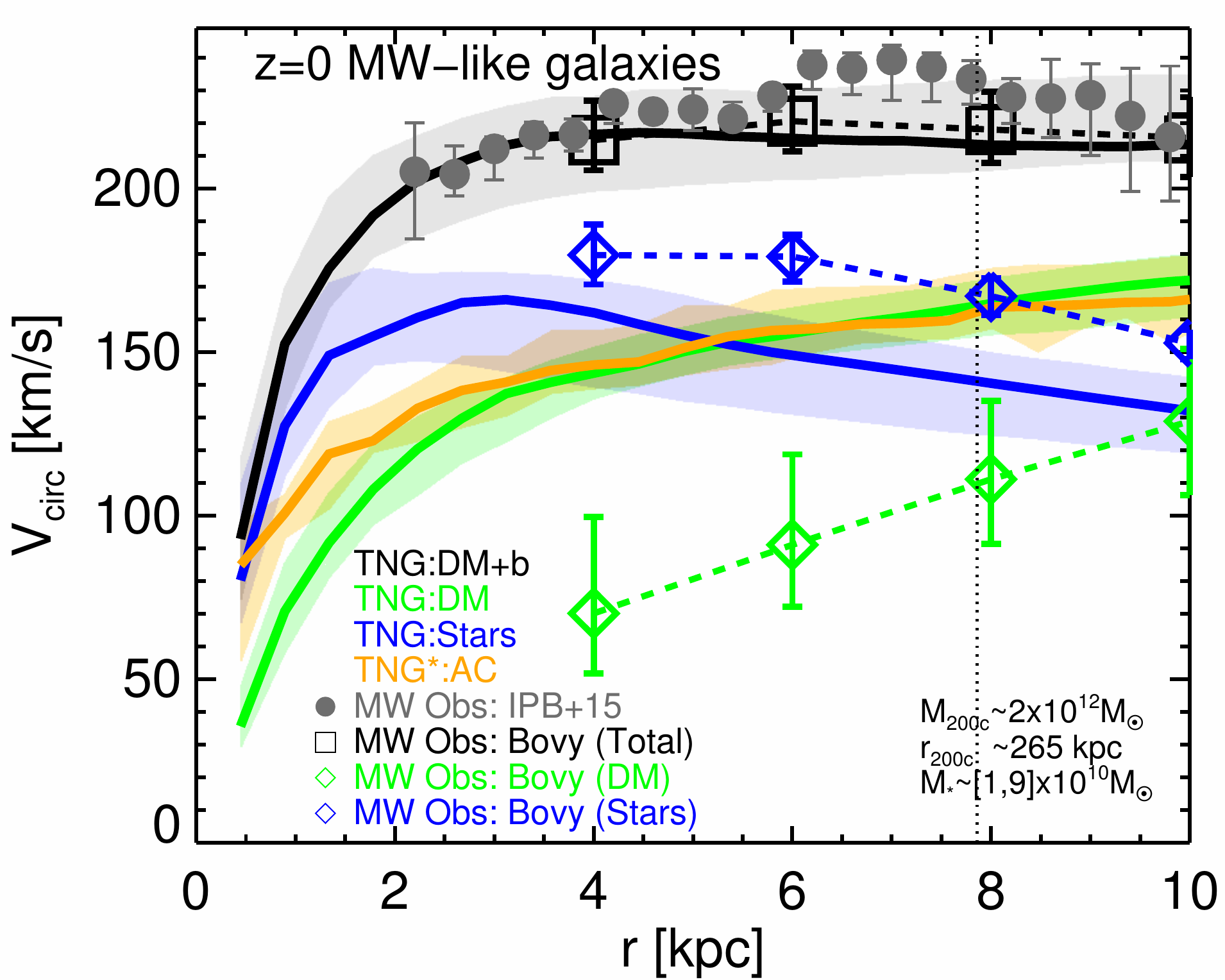}%
 \caption{Median circular velocity curves of 20 TNG haloes with $M_\rmn{200c}\sim 2\times10^{12}\Msun$ that host disc-like galaxies. These are the galaxies  shown in Fig.~\ref{BOVY}, bottom left panel. In addition to the data shown in that Figure, we here include the median expected dark matter distribution under adiabatic contraction (orange line), median stellar mass component from the simulations (solid blue line) and that derived for the stellar disc by \citealt{Bovy13}. The remainder of the thick lines indicate median TNG predictions and the dashed lines / symbols show observational results: green for dark matter and black for DM+baryons. The 68~per~cent region for our sample of 20 galaxies is shown as a series of coloured bands, one each for DM+b, dark matter, and stellar mass. The horizontal dashed line indicates the median $r_\rmn{half}$ of the TNG galaxy sample.}
    	\label{BOVY2}
\end{figure*}

Starting with the total mass (DM+b, black curves), the circular velocity profiles of TNG MW-like galaxies exhibit shapes and degrees of flatness within about 10 kpc from the centres in good agreement with the Galaxy rotation curve inferred by \citealt{Bovy13,IPB15}. In contrast, the total mass profiles of MW-like galaxies from the Illustris simulation rise too steeply with radius, reflecting the aforementioned problems in the galaxy stellar sizes and hence being ruled out at all MW-like mass scales.

Based on the total matter distribution alone, the TNG model would favour a large Galaxy halo mass, of the order of $1.5-2 \times 10^{12}\Msun$. On the other hand, in TNG we consistently obtain higher dark matter circular velocity curves (green lines) than the best-fit dark matter component of the \mbox{\citet{Bovy13}} dynamical model (see open green diamonds). In this regard, we obtain a similar result to the EAGLE and Auriga hydrodynamical simulations \citep{Schaller15,Grand17}, with the latter predicting $V_\rmn{c,DM}(6~\rmn{kpc})>150$~\kms for simulated galaxies that match the Galaxy's rotation curve.  Only the dark matter mass profiles from the DMO analogues of TNG galaxies (grey curves, especially at the low-mass end) seem to approach the constraints from \citet{Bovy13}, although being consistently flatter than the inferred ones for the Galaxy. So for example, for the $8\times10^{11}\Msun$ halo mass bin, the average TNG halo could host the Galaxy rotation curve if i) the stellar masses were higher than TNG100 predicts -- for which there is room, since the highest stellar mass of the 251 disc-dominated galaxies within 10~per~cent of $M_\rmn{200c}=8\times10^{11}\Msun$ is $3.6\times10^{10}\Msun$ and thus less massive than the Galaxy estimate of $4-6\times10^{10}\Msun$ \citep{McMillan11,Bovy13} and ii) if the contraction of the halo could be avoided. 

The dark matter circular velocity curves in TNG are consistent with a dark matter density profile that has an average power-law slope of about $-1.8$ in the $5-10$ kpc radial range (purple dots): we suggest that such an alternative functional form could be allowed when modelling the dark matter contribution in observations -- although this slope does appear at the current time to be steeper than is permitted by the microlensing study of \citet{Wegg16}, which prefers an NFW profile.

We emphasise the preference for a more massive halo, and the apparent tension between observations and our results, in Fig.~\ref{BOVY2}, which includes also the predicted median {\it stellar} circular velocity profile (blue curves) in addition to that derived for the Galactic stellar disc by \citet{Bovy13}, and the adiabatic contraction prediction for  TNG galaxies as calculated using \citet{Blumenthal86} and \citet[][orange curve]{Gnedin04}. The slopes of both the stellar and dark matter components have the same sign as that of the observations, in that the stellar component is falling and the dark matter component rising. The amplitudes of the two curves are instead very different: the simulations predict that the MW is baryon-dominated only within 6~kpc, as does the adiabatic contraction model, which agrees remarkably well with the simulation result over the radius range probed by \citet{Bovy13}; the \citet{Bovy13} study instead strongly prefers that dark matter dominate the mass budget only outside 10~kpc. We anticipate that a crucial test of all future galaxy formation models will be whether they can achieve simultaneously a sufficiently massive and compact MW galaxy without over-contracting its host halo.

In closing this comparison to the mass distribution in our Galaxy, it is important to keep in mind, however, that there is still some considerable divergence in the inferred dark matter fractions across different observational studies. In fact, while all observational results from dynamical models agree on the distribution of the {\it total} matter, the separate contributions of dark and baryonic components are highly uncertain, with \citet{IPB15} reporting $f_{\rm DM} (\la 10 {\rm  kpc})<0.6$, \citet{Bovy13} favouring $f_{\rm DM} \sim 0.3$ at 2.2$~R_\rmn{d}\approx4.7$~kpc, and other independent estimates from microlensing studies reading $f_\rmn{DM} (\la 10 {\rm  kpc}) <0.5$ \citep[e.g.][]{Hamadache06,Wegg16}. Together with other analyses that estimate the dark matter fraction at about 45 per cent at galactocentric distances of $4.5 - 12$ kpc \citep[][see Table~\ref{ObsTab}]{BlandHawthorn16}, this observational landscape places the average DMFs predicted for MW-like galaxies by the TNG and Illustris model a factor $\sim 1.2-3.0$ higher than observational constraints for our Galaxy when measured within 5~kpc. In fact, we have checked that among the many hundreds of MW-like galaxies in the TNG100 box, {\it none} of them can simultaneously satisfy the circular velocity profiles of both the total and the dark matter components derived e.g. by \mbox{\citet{Bovy13}} and reported in Fig.~\ref{BOVY}. 

We therefore conclude that either (i) our model may be missing some important physical aspects, (ii) this discrepancy may reflect differences between our theoretical circular velocity curves and the observationally determined rotation curves, or (iii) some combination thereof. In this case, careful virtual observations may provide a resolution particularly in the inner regions of galaxies where peaks in the rotation curve can be washed out observationally. Such a study will also clarify the uncertainty in the total mass of faint stars, which is degenerate with the mass in dark matter \citep[e.g.][]{Smith15} , and will shed light on the validity or limitations of the assumptions that are typically adopted in the mass modelling methods, which in turn are used to translate observational data into mass profiles.

\subsection{Low redshift bulge-dominated galaxies}

We now transition our focus to low-redshift galaxies that are broadly classified as ellipticals or early type galaxies (ETGs), or, for our methodology, bulge-dominated. 

\begin{figure*}
 \includegraphics[width=8.2cm]{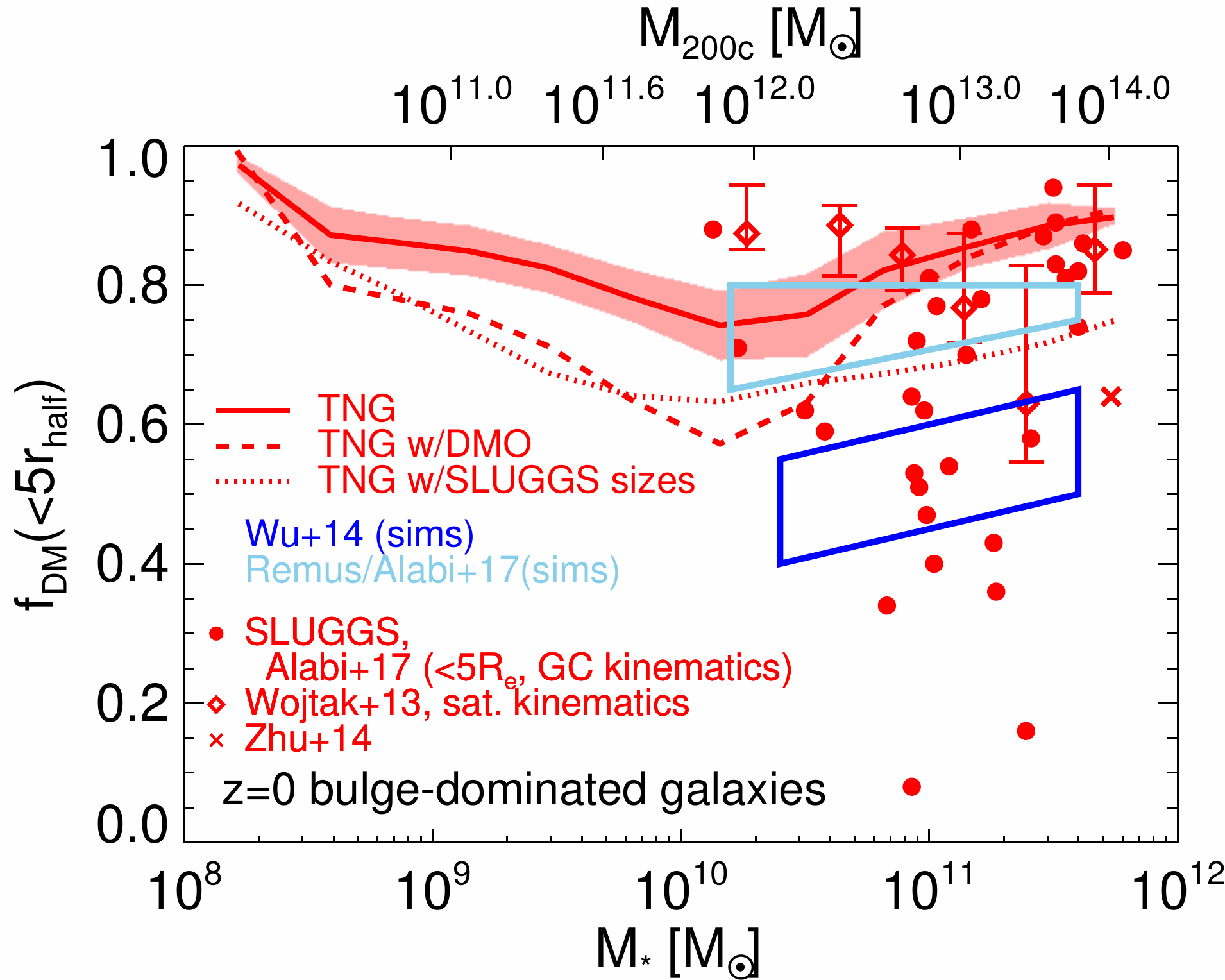}
  \includegraphics[width=8.2cm]{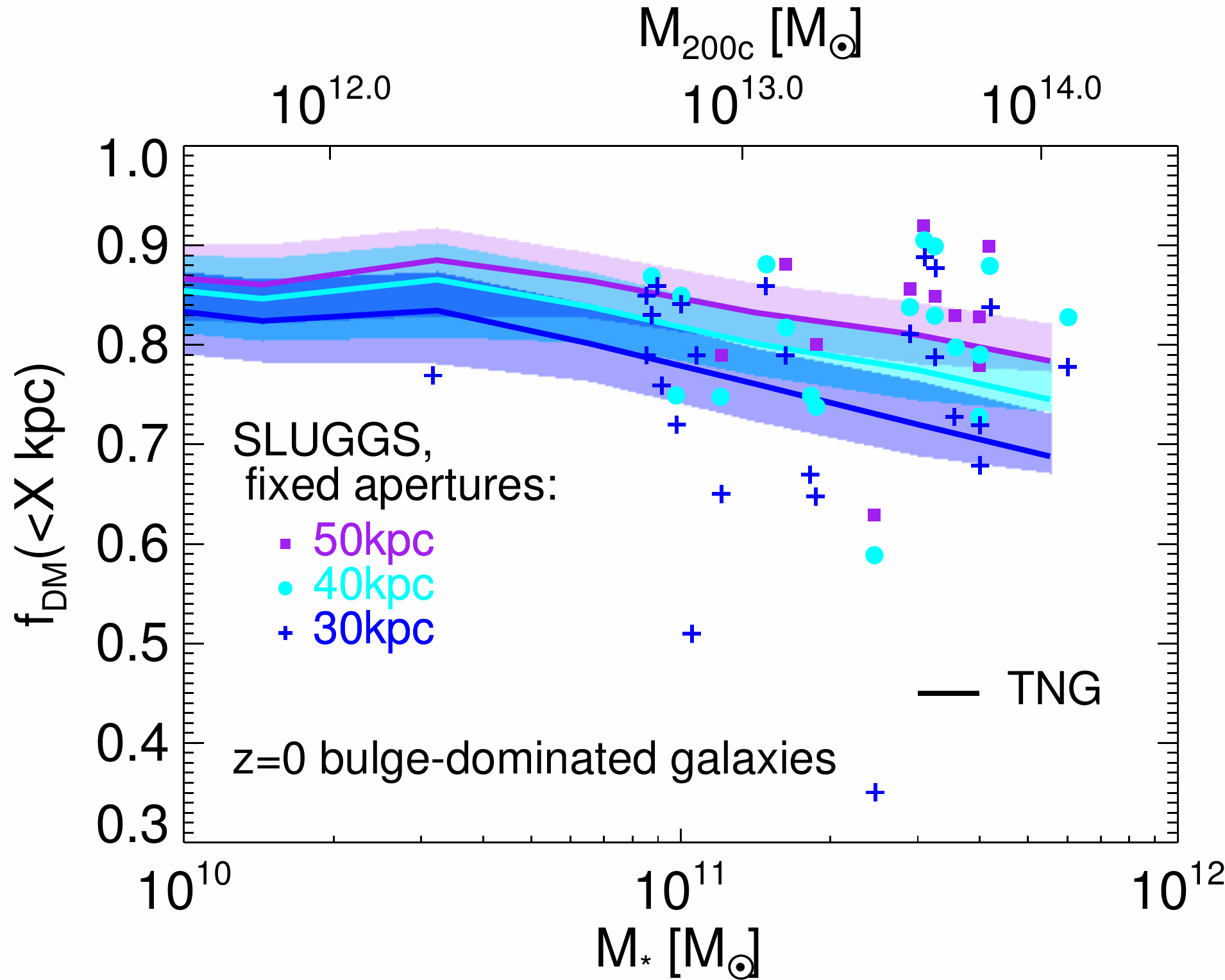}\\
  \includegraphics[width=8.2cm]{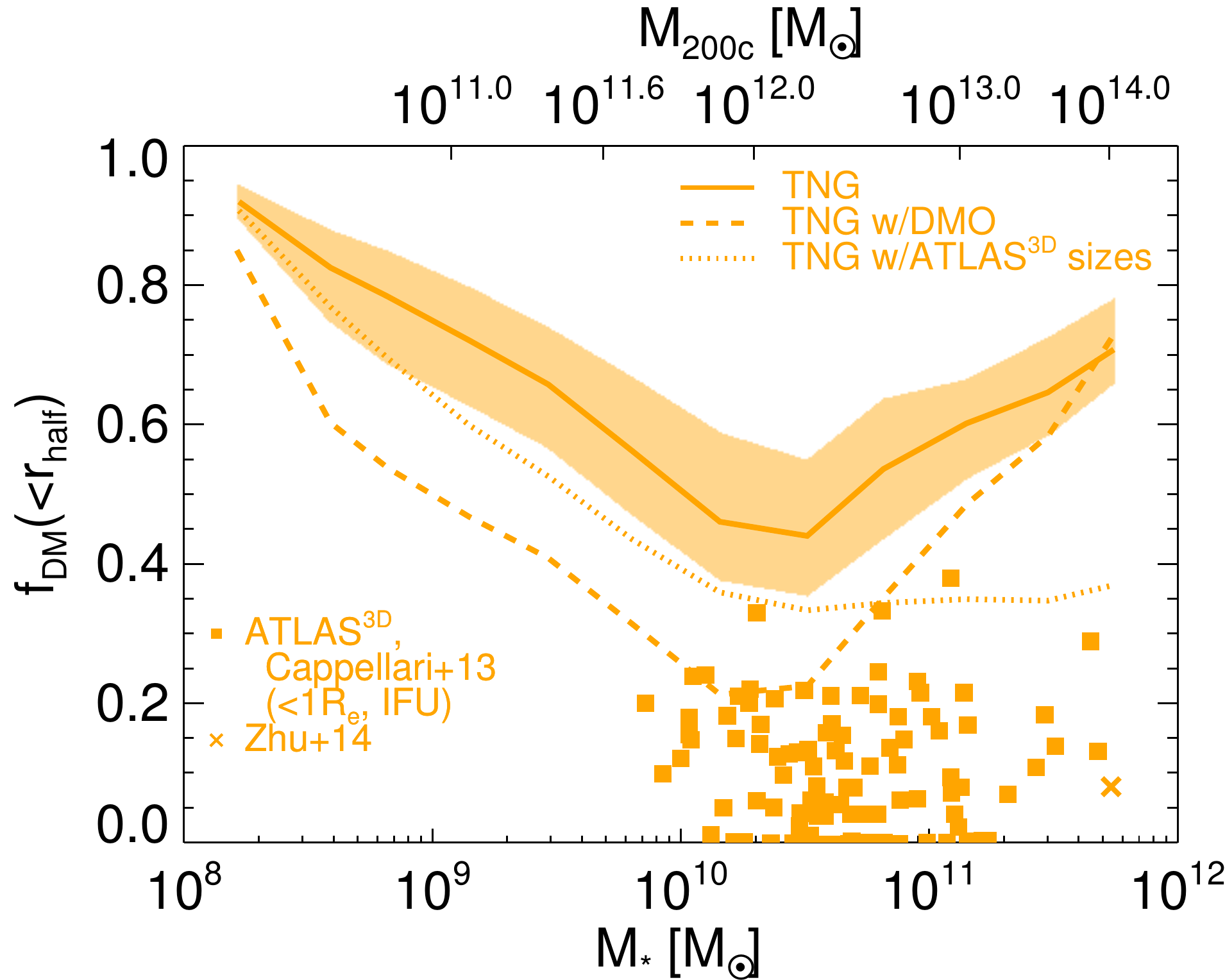}
   \includegraphics[width=8.2cm]{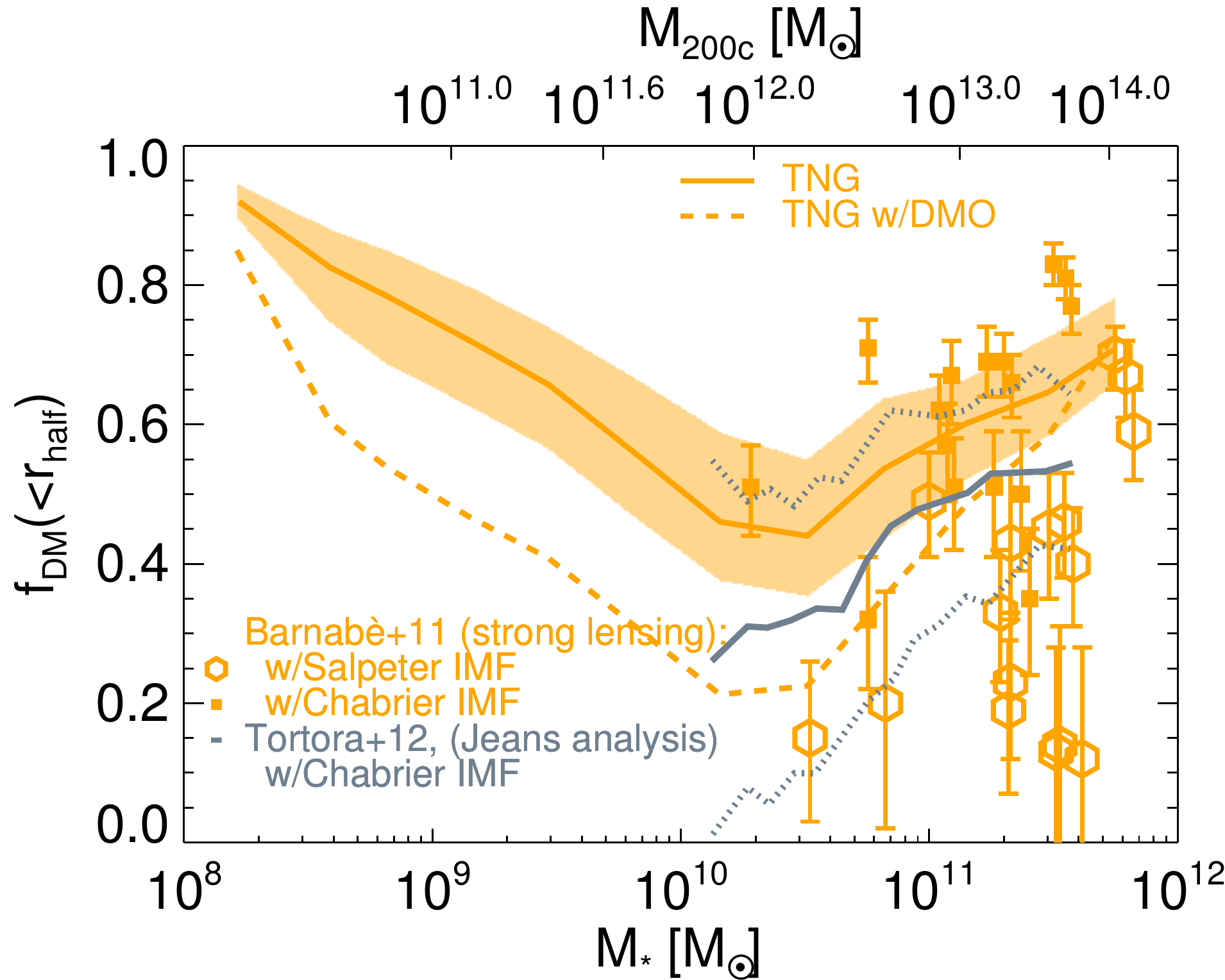}
 \caption{ Dark matter fractions of bulge-dominated galaxies as a function of stellar mass at $z=0$. The top left panel shows the dark matter fraction measured within 5 $r_\rmn{half}$ radii, the top right panel within a series of fixed apertures, and the bottom two panels within 1 $r_\rmn{half}$. Each panel includes observational data obtained by a different method. In the top left panel, the method is globular cluster and satellite kinematics (\citealp{Alabi17}, filled circles; \citealp{Wojtak13}, open circles), and we also include the regions of parameter space occupied by the simulation results of \citet{WuX14} (dark blue) and \citet{Remus17} (light blue, adjusted to 5 $r_\rmn{half}$ instead of 1 $r_\rmn{half}$, as indicated in \citealt{Alabi17}). In the top right panel we show the results obtained from SLUGGS when three fixed radius apertures are used (for those galaxies where that is possible): 30~kpc (blue crosses), 40~kpc (cyan circles) and 50~kpc (purple squares). The corresponding TNG data is shown in solid lines, with colours that match the three apertures. We also shrink both axes of this panel for clarity. In the bottom left panel we show the results the IFU study of \citet{Cappellari13} (squares). The M87 study by \citet{Zhu14} uses both IFU and GC orbits and therefore features as a cross in both the top left and bottom left panels. In the bottom right panel we include gravitational lensing and Jeans modelling studies, and also show the crucial uncertainty introduced by the IMF. The data of \citet{Barnabe11} are shown as squares (with the Chabrier IMF) and open hexagons (Salpeter IMF); the grey line shows the median DMF-$M_{*}$ relation of \citet{Tortora12} assuming a Chabrier IMF, and the dotted grey lines the upper and lower quartiles. The simulation results (TNG100) are given as solid curves with one-sigma galaxy-to-galaxy variations denoted by shaded area. In all panels, dashed curves denote the results we would obtain by placing TNG galaxies in NFW (dark matter only) dark matter haloes, i.e. by neglecting the dark matter contraction predicted by the full-physics TNG model. Dotted curves indicate the DMFs measured when the galaxy sizes are drawn from the median stellar mass-size relations of the SLUGGS (top-left panel) and ATLAS$^\rmn{3D}$ (bottom-left panel) surveys.}
 
    	\label{SLUGGS}
\end{figure*}

 Recently, within the {\sc sluggs} survey, the dark matter content of a set of bright elliptical galaxies in the Local Universe has been inferred from the orbits of nearby globular clusters \mbox{\citep{Brodie14,Alabi16}}. The quoted dark matter fractions have been measured within 5 effective radii, the latter being the half-light radius of a galaxy seen in projection in the sky \citep{Alabi17}. They find a wide variety of DMFs in their sample, from as high as 0.95 to as little as 0.05, and therefore challenge theory to explain both the median DMF and also a very large scatter. These values are reported in Fig.~\ref{SLUGGS}, upper left panel. There we also include data from a complementary study by \mbox{\citet{Wojtak13}}, who use satellite galaxies instead of globular clusters as mass tracers to obtain dark matter fractions within the same 5 effective radii aperture. The latter adopts a Chabrier IMF \citep{Chabrier03} not dissimilar to the Kroupa IMF used by \citet{Alabi17}. \citet{Wojtak13} find a smaller range of dark matter fraction values for ETGs, which are always larger than 50~per~cent.

 Previous constraints had been provided with the {\sc atlas$^\rmn{3D}$} survey, in which the DMF of local early-types had been measured within instead just one effective radius \mbox{\citep{Cappellari13}}. These measurements prefer very low DMF at small radii, almost completely below 0.2 (see Fig.~\ref{SLUGGS}, lower left).  This may be in part due to their preferred IMF, which transitions from a Kroupa IMF at low velocity dispersions to a Salpeter form \citep{Salpeter55} for massive ETGs \citep{Cappellari13b}. The difference made by the choice of the underlying IMF has been explicitly shown by \citet{Barnabe11} who repeated their analysis with both the Salpeter and Chabrier IMF as part of a lensing study (see Fig.~\ref{SLUGGS}, lower right). The values of DMFs that rely on or imply a Chabrier IMF are systematically larger by about 40~per~cent (absolute percentage) than those obtained adopting a Salpeter-like form. A similar set of results was found by \citet{Tortora12}, who instead combined a Chabrier IMF with a Jeans analysis (Fig.~\ref{SLUGGS}, lower right). 
In Fig.~\ref{SLUGGS}, large crosses are the DMFs obtained for M87 by \citet{Zhu14} for 5 (top panel) and 1 (lower left panel) effective radii, respectively; we include this study as an example of a paper that focuses on one single galaxy and whose subject was analysed separately for SLUGGS. 

Most of the studies above -- with the one exception of the \citet{Barnabe11} lensing study -- provide the dark matter fraction within projected 2D apertures. We instead measure our sizes in 3D stellar half-mass radii rather than in projected half-light radii, since the former are more important physically and are also more easily measured in simulations. It has been shown that the 3D mass-based and 2D r-band based sizes of TNG galaxies are essentially identical to one another for stellar masses of $\sim 10^{10.5} \Msun$ and above \citep[][Fig. 2]{Genel17}. This points towards some relevant issues relating to the simulated and observationally-adopted values and radial trends of the stellar mass to stellar light ratios. For the purposes at hand, this fact allows us to show the 3D-aperture DMFs of our simulations directly against those of the observational 2D apertures, e.g. \citet{Alabi17} within 5 effective radii.

In Fig.~\ref{SLUGGS}, solid curves denote the TNG running averages equipped with the one sigma galaxy-to-galaxy variations. Dashed curves denote the TNG DMFs we would obtain by placing TNG galaxies in NFW DM haloes, i.e. by neglecting the dark matter contraction emerging from our full-physics model. 
All choices of aperture show well defined relations for TNG galaxies: the DMFs increase systematically with stellar mass at $M_{200c}>3\times10^{12}\Msun$. Moreover, the smaller aperture predicts lower DMFs than the larger one, yet in the range of $f_{\rm DM} (< 2-5 \times r_{\rm half}) \sim 0.5-0.9$. 

We remind the reader that TNG galaxies have been modelled adopting a Chabrier IMF and could not be hence directly compared to observational results that seem to favour or assume a very different IMF shape. In the upper left panel of Fig.~\ref{SLUGGS}, the width of the simulated distributions (10 per cent at 1$-\sigma$) is much smaller than the spread of the {\sc sluggs} data (40 per cent, accounting for both an intrinsic scatter and a measurement contribution), but are broadly consistent with the findings of \citet{Wojtak13} (filled circles and open circles respectively) although the latter prefers DMFs at $2\times10^{10}\Msun$ significantly higher than those of TNG. We also include a broad reference to two previous simulation studies, \citet{WuX14} and \citet{Remus17}. The latter paper produces galaxies with DMFs qualitatively similar to ours at the lower end of the displayed mass range masses, but with lower fractions at higher masses; the \mbox{\citet{WuX14}} paper instead predicts consistently lower DMFs, which is likely due in part to the absence of AGN feedback given the extra star formation and the further halo contraction shown above. Crucially, none of the models matches the spread and median DMF values of the {\sc sluggs} data, even when we allow for the possibility of no halo contraction by placing TNG100 galaxies in their DMO counterparts (dashed curves vs. both sets of circles). However, we also recognise a disagreement between different observational measurements: \citet{Zhu14} measured a 5 effective radii DMF of 0.63 for M87, whereas SLUGGS instead found a DMF of 0.86 for the same galaxy. 

The  apparent discrepancy is even more acute for the intra-galactic measurements by \citealt{Cappellari13} (filled squares, bottom left panel), which determine lower DMFs than even the non-contracted TNG galaxies with $1\times r_\rmn{half}$ (dashed yellow curve). Their quoted DMFs are obtained by imposing an NFW dark matter profile and are very sensitive to the choice of dark matter slope, and a steeper slope -- such as that produced in our simulations -- may instead lead to better agreement with our results. However, a recent study by \citet{Wasserman17} of the matter distribution in one of the {\sc sluggs} galaxies, NGC~1407 with estimated stellar mass of $\sim2\times10^{11}\Msun$, argued for a dark matter slope no steeper than $-1$. 
 Importantly, the intra-galactic DMFs of the lower left panel are more than a factor of two smaller than those obtained by \citealt{Barnabe11} and \citealt{Tortora12} (lower right, filled squared and grey curve vs. yellow solid curve). When a consistent choice for the IMF is adopted in both simulations and observations (i.e. Chabrier), the latter observational results are in good agreement and at most 10~per~cent lower than the TNG predictions, respectively. By contrast, the Salpeter IMF choices consistently imply much lower DMFs, and a variable IMF would bring the \citet{Tortora12} result into better agreement with that in ATLAS$^\rmn{3D}$ \citep{Tortora13}. Note that our simulations were performed with the Chabrier IMF; if we were to instead adopt a Salpeter IMF, the available energy from supernova feedback would decrease, leading to higher stellar masses and correspondingly lower DMFs, but for the role of the AGN. 

In fact, a number of assumptions are required in the mass modelling that leads to the observational constraints on the DMFs reported thus far. These include not only the already-discussed choice for the IMF, but also assumptions for the shape of the total potential, the value and radial dependence of the orbital anisotropy, and the value and radial dependence of the stellar mass to stellar light ratios. So, for example, the SLUGGS and \citet{Zhu14} data assume spatially-constant stellar mass to light (M/L) ratios in the underlying dynamical modelling: relaxing such assumptions and allowing for some M/L gradients across the observed galaxies may instead return larger inferred DMFs than those reported in Fig.~\ref{SLUGGS}, possibly bringing simulations and observations into better agreement.
 One such example of the adoption of a M/L gradient is \citet{Tortora12}, who applied a single isothermal sphere model of M/L together with a Chabrier IMF. They obtain a median relation similar in shape and gradient to that of the TNG galaxies, although 0.1 points lower. The scatter is also 1.5 times larger than is the case for the simulations. Interestingly, the \citet{Tortora12} median is located above the DMO expectation value, indicating that their data does prefer some contraction of the host dark matter halo, although not as much as we predict.

 Finally, we note that the galaxy apertures adopted for the {\sc sluggs} and {\sc atlas$^\rmn{3D}$} results are not consistent with those used for the TNG DMFs. For galaxy masses of $\sim 10^{10.8} \Msun$ and above, the TNG galaxy sizes (3D or 2D) are up to a factor 2.5 larger than those measured in {\sc sluggs} results (see Table 1 of \citealt{Alabi17}), even when marginalizing over the 0.1-0.2 dex uncertainties that are intrinsic to any galaxy stellar-mass estimate. Even more crucially, the galaxy-size values derived from the {\sc atlas$^\rmn{3D}$} galaxies that are then used for the {\sc atlas$^\rmn{3D}$} DMFs are, in turn, also a factor of a few 10s of per~cent lower than those estimated for the most massive {\sc sluggs} galaxies, at least at $M_{\rm stars} \sim 10^{10.11} \Msun$ and above. As the mass fraction in dark matter has a very steep radial profile, especially within the innermost regions of galaxies (see Fig.~\ref{VcircM}, right column), it is crucial to evaluate comparisons of DMFs at consistent apertures. We can quantify the effects of the mismatched apertures and account for this in our simulated measurements. In this exercise, we recognise the possibility that the simulated and observed galaxy structures may be similar but the methods to measure effective radii may be incompatible, and hence produce inconsistent galaxy-size labels. In Fig.\ref{SLUGGS}, dotted curves denote the TNG galaxy DMFs evaluated within apertures that correspond, not to the values of the TNG stellar half-mass radii, but to those adopted in the {\sc sluggs} (top panel) and {\sc atlas$^\rmn{3D}$} (bottom left panel) data. These aperture-corrected DMFs from the TNG simulated galaxies are generally 15 and 20~per~cent lower than those evaluated at larger apertures for the $5\times r_\rmn{half}$ and $1\times r_\rmn{half}$ apertures respectively. This brings the median of the TNG data to be fully consistent with the median of the SLUGGS data (top left, 15 of the 32 SLUGGS points lie below the dotted line), but it only mildly reduce the tension with the ATLAS$^\rmn{3D}$ estimates. This residual discrepancy can have three possible causes. First, if the sizes of TNG galaxies are much larger than the true sizes of ATLAS$^\rmn{3D}$  galaxies, much of the TNG stellar mass lies at large radii and so the central DMF is still higher than ATLAS$^\rmn{3D}$ expects. Second, TNG predicts too much dark matter contraction in the innermost regions of galaxies, and thirdly the disagreement between \citet{Cappellari13} and \citet{Zhu14} on the one hand and \citet{Barnabe11} and \citet{Tortora12} on the other may indicate that future analyses of the ATLAS$^{3D}$ galaxies may return a different answer.

Alternatively, we can calculate the observational and simulated galaxy DMFs within {\it fixed} apertures to lower the uncertainty due to the simulated galaxy sizes. We finally plot the DMFs measured for TNG ellipticals and SLUGGS galaxies in 50, 40 and 30~kpc apertures in the upper-right panel of Fig.~\ref{SLUGGS}. The scatter decreases in both datasets and the centroids of the distributions agree much better than within the $5\times r_\rmn{half}$ aperture. For the most massive galaxies the simulations actually underestimate the DMFs slightly, but this discrepancy is small in comparison to the case when using multiples of $r_\rmn{half}$.

We conclude that the TNG simulations do a reasonable job of matching observational findings that assume a Chabrier IMF i.e. an IMF choice consistent with the one adopted in the simulations, and that the apparent inconsistency with e.g. the {\sc sluggs} (but not the {\sc atlas$^\rmn{3D}$}) results can be fully alleviated by measuring the DMFs within consistent aperture values. Those uncertainties in the IMF plus the source of scatter are remaining issues to be addressed in future work. We advocate for future results to be expressed in terms of DMFs within fixed physical apertures in kpc (e.g. 5, 10, 25 or even 50~kpc, see Fig.~\ref{fDM}, or the whole profiles, see Fig.~\ref{VcircM}) even if across a wide range of galaxy masses. This approach would at least short circuit the additional source of systematic uncertainties that are associated to the measurements of the galaxy sizes.

\begin{figure*}
 \includegraphics[width=8.8cm]{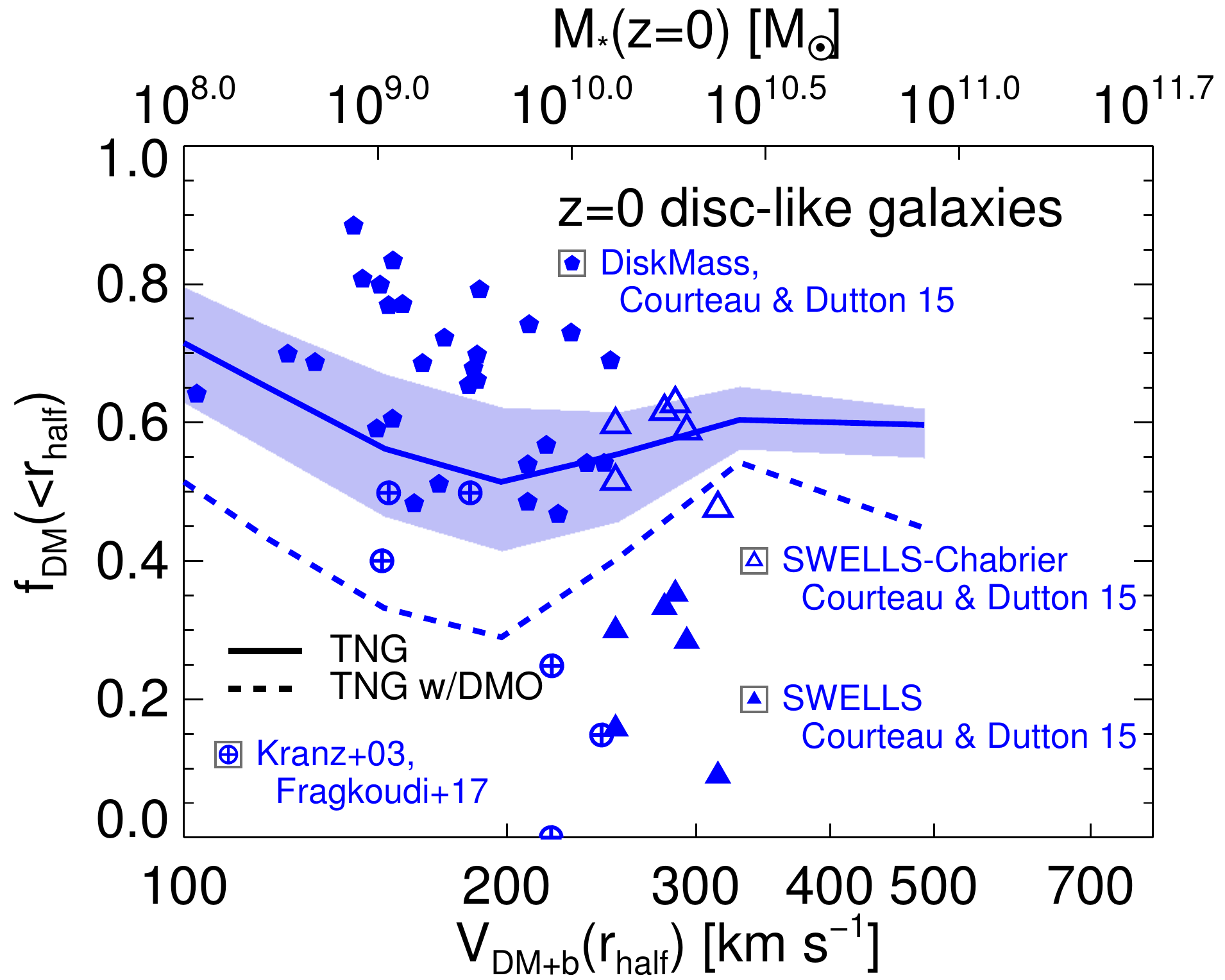}
 \includegraphics[width=8.8cm]{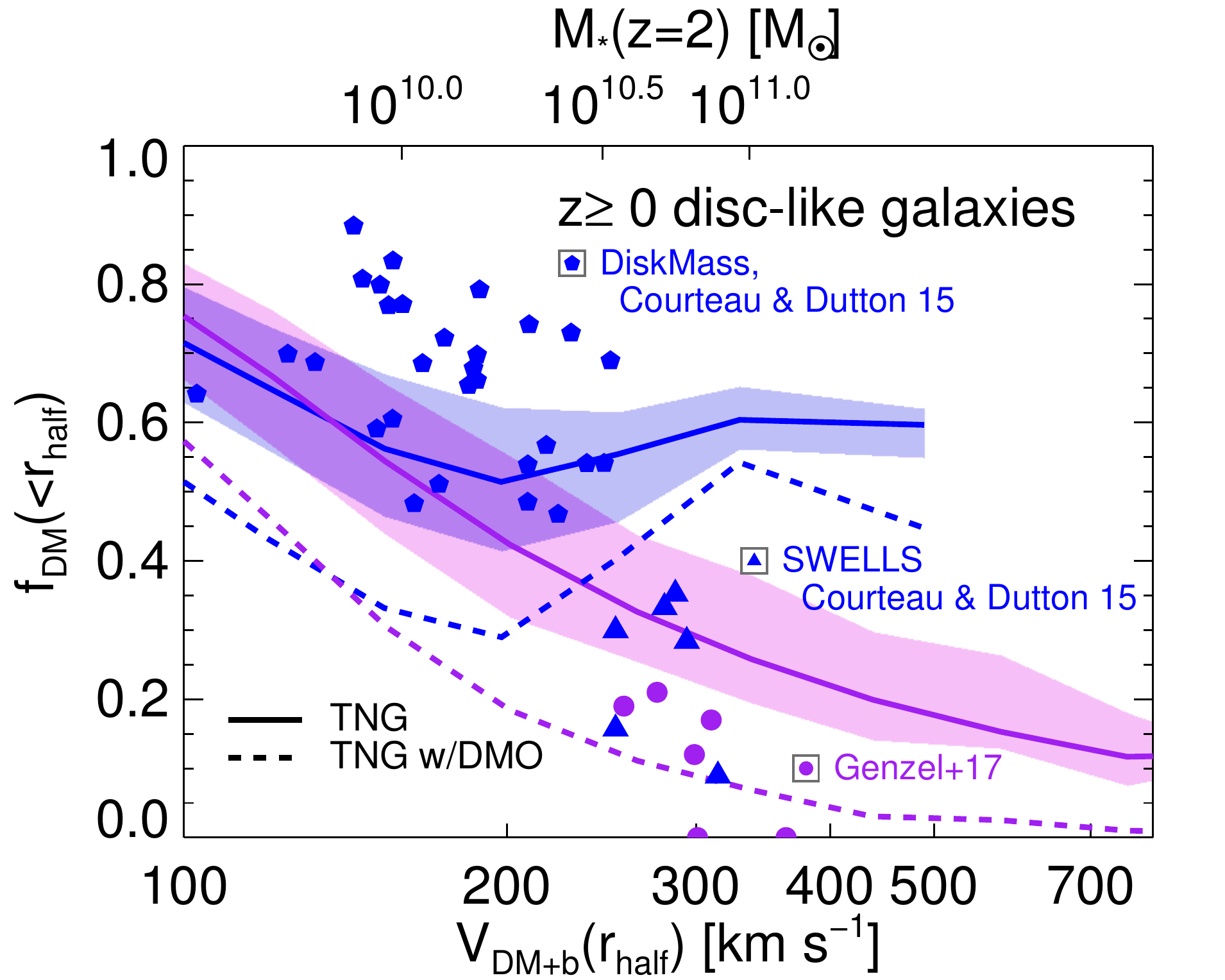}
 
\caption{DMF of TNG100 disc-dominated galaxies as a function of the circular velocity measured at $r_\rmn{half}$ (3D).  Left-hand panel: the DMF of TNG $z=0$ discs with the observational results that \citet{Courteau15} derived for the SWELLS \citep{Barnabe12,Dutton13} and DiskMass \citep{Martinsson13b,Martinsson13a} galaxy samples included as blue triangles and pentagons respectively. The Disc DMF measurements of \citet{Kranz03} and \citet{Fragkoudi17} are shown as crossed circles. We also include a version of the SWELLS data in which we change the IMF \emph{a posteriori} from Salpeter to Chabrier (empty triangles), to make it consistent with our simulation choice. The error bars on the data points have been omitted for clarity: they are typically of order 0.1~units in DMF. 
Right hand panel: DMFs for galaxies at $z=0$ (blue) and $z=2$ (purple), plus some of the observations presented in fig.~3 of \citet{Genzel17}. We include the results of \citet{Genzel17} as purple circles. In both panels, solid curves are the TNG running medians and dashed lines denote DMO-counterpart medians.}
    	\label{GENZEL}
\end{figure*}

\subsection{Disc-dominated galaxies across times}
\label{sec:Ddgat}
Our last observational comparison focuses on disc galaxies across masses and times:  Fig.~\ref{GENZEL}. Observations seem to suggest smaller DMFs at higher redshifts: for $z=2$ discs, \citet{Genzel17} obtain average DMFs within 1 effective radius of just 0.15 (dark grey filled circles). This is lower than disc galaxies in the Local Universe of similar masses, although measured at larger apertures: DiskMass galaxies (light grey pentagons) exhibit DMFs in the range $0.5-0.9$ within 2.2 times the disc scale radius, $R_\rmn{d}$, (or approximately 1.6 times the 2D effective radius) \mbox{\citep{Martinsson13a,Martinsson13b}}, and at similar redshifts and apertures the SWELLS galaxies \mbox{\citep{Barnabe12,Dutton13,Courteau15}}, which have larger bulge components than  the DiskMass galaxies, seem to favour lower values, $f_{\rm DM} \sim 0.1-0.4$ \citep{Courteau15} at least in part because their bulge components prefer a Salpeter IMF to a Chabrier. From a theoretical perspective, we have already shown in Fig.~\ref{RHvZPlus} that the DMF at high galaxy masses can change considerably with redshift due to the increase in the sizes of galaxies at fixed stellar mass. 

We now apply this result to the specific case of disc-dominated TNG100 galaxies at $z=2$ (violet curves) and at $z=0$ (blue curves) in Fig.~\ref{GENZEL}, by plotting the DFMs within the simulated 3D stellar half mass radii\footnote{Note that $2.2R_\rmn{d}$ is slightly larger than one time the 3D stellar half mass radii. Given the diversity of DMF values reported by the various observational data sets, here we simplify the comparison by allowing systematic discrepancies of $\sim$~10~per~cent across studies.}. We also show the  \citet{Genzel17}  data, the  \mbox{\citet{Courteau15}} DMF derivations for $z=0$ galaxies, a modified version of the \citet{Courteau15} SWELLS derivations that uses the Chabrier IMF, and a set of disc observations by \citet{Kranz03} and \citet{Fragkoudi17} that assume different systematics. The data are separated into two panels: the left-hand panel explores some of the systematic uncertainties in the $z=0$ observations, and the right-hand panel shows the difference between $z=0$ and $z=2$.

In the left-hand panel of Fig.~\ref{GENZEL} we look at $z=0$ TNG galaxies and their systematics.  At higher velocities/masses, the $z=0$ DMF increases slightly, and in the median is systematically $\sim15$~per~cent lower than the centre of the \mbox{\citet{Courteau15}} DiskMass results but considerably larger than the $z=0$ SWELLS results.\footnote{Part of the discrepancy with the DiskMass galaxies is due to $2.2R_\rmn{d}$ being 58~per~cent larger than the effective radius. However, since our $M_{*}<10^{10.5}\Msun$ galaxies have 3D apertures 46~per~cent larger the 2D effective radii, our apertures are only $\sim10$~per~cent smaller than those of DiskMass and thus only slightly underestimate the DiskMass dark matter fractions.} We show the importance of the choice of IMF by calculating the DMFs that the SWELLS galaxies would have if the bulge IMF were Chabrier rather than Salpeter. We do this by dividing the stellar mass within $r_\rmn{half}$ by a Chabrier-Salpeter conversion factor of $10^{0.20}$, keeping the total mass within $r_\rmn{half}$ constant and then adjusting the dark matter mass accordingly (we assume that the gas contribution is negligible and that the stellar mass is dominated at this radius by the Salpeter bulge rather than the Chabrier disc). This correction brings these galaxies into good agreement with the TNG simulations. We also show the results for a series of measurements by \citet{Kranz03} and \citet{Fragkoudi17}. Their discs show consistently lower dark matter fractions than the DiskMass galaxies of the same mass, and \citet{Fragkoudi17} states that this is due to a different choice of tracer for measuring the stellar disc mass. Further, virtual observations will be required to check whether these systematics can explain both the discrepancy between our results and observations plus the differences between the observations themselves.

In the right-hand panel of Fig.~\ref{GENZEL} we compare our $z=0$ DMFs to those measured at $z=2$. In TNG, the DMF at fixed circular velocity is roughly invariant across redshifts in the 100-200~\kms range, and as stated above $z=0$ galaxies exhibit a small increase in DMF for circular velocities $>200$~\kms. The higher redshift simulated galaxies instead continue to still lower DMFs in a power-law-like fashion. This change is much more in keeping with the low DMF data of \citet{Genzel17}, that are however shifted by $\sim 0.15$ towards lower DMFs than the average TNG galaxies. It is interesting to note that TNG galaxies in NFW, un-contracted halo  counterparts (dashed violet curve) give a remarkably good agreement with the $z=2$ data but possibly \emph{worse} agreement with the present day discs (e.g. dashed blue curve vs. blue pentagons), which are more extended. In the models, both redshifts return up to a factor of two lower DMFs when the DMO dark matter masses are used. It may therefore be the case that an excess of contraction in the inner parts of the halo at high redshifts is also part of the reason for the disagreement. It will be interesting to see in the coming years how different simulations exhibit different degrees of halo contraction across galaxy masses and redshifts, and what processes are most effective at mitigating the effects thereof. In fact, the separation between the observationally-inferred and simulated DMFs is sufficiently small that a more careful treatment of the simulations to produce virtual observations including separate treatment of gaseous and stellar tracers and inclination effects, may in fact bring good agreement with these $z=2$ discs. If the errors on the observations contain a systematic uncertainty, this boost of 10~per~cent to the observed DMF will also play a role in bringing better agreement.

\begin{figure}       
        \includegraphics[scale=0.65]{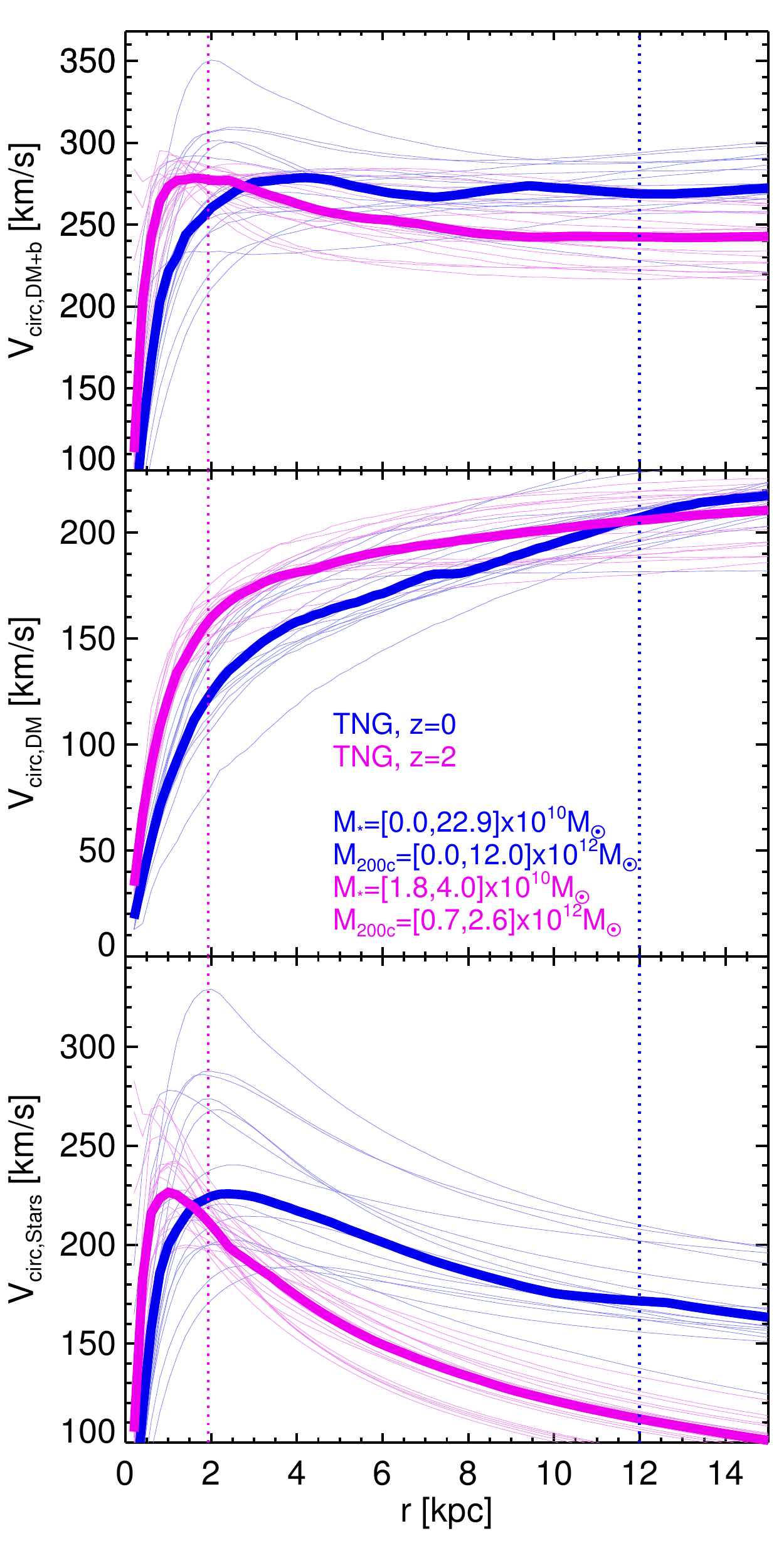}
        \caption{Circular velocity profiles for the total (top row), dark matter (middle row), and stellar (bottom row) components of a sample of 20 $z=0$ and $z=2$ TNG100 central disc-dominated galaxies with circular velocity at $r_\rmn{half}$ of about 280~\kms. The $z=0$ data are shown in blue and the $z=2$ data in purple. The two horizontal lines show the the median $r_\rmn{half}$  of the two samples. The range of stellar and halo mass in each sample is presented in the top panel inset.}
        \label{z0vz2}
\end{figure}

Finally, we have also checked the gradients of our $z=2$ circular velocity curves. We show a sample of 20 $z=2$ disc-dominated galaxies in Fig.~\ref{z0vz2}, which have been selected to have a circular velocity at $r_\rmn{half}$, $V_\rmn{c,DM+b}(r_\rmn{half})$, of around 280~\kms. We include panels for the total mass, dark matter and stellar mass components, along with 20 $z=0$ disc-dominated galaxies that have the same approximate 280~\kms, although such high circular velocity disc galaxies are rare at $z=0$ and so in practice the highest $V_\rmn{c,DM+b}(r_\rmn{half})$ in our sample is $\sim320$~kpc. 

We find that about half of these $z=2$ median circular velocity curves are decreasing in amplitude at the sample's median half stellar mass radius, which is actually not too different to the $z=0$ sample in the literal sense of a `falling' circular velocity curve (top panel). However, the {\it median} circular velocity curve is peaked around the median stellar mass radius, and falls thereafter, whereas in the $z=0$ case the curve at the median stellar mass radius is instead flat (top panel). This is due to the higher concentration of the baryonic component at high redshifts, in a way that is not reproduced at lower redshifts (bottom panel), and is in spite of the higher concentration of the dark matter component at high redshift (middle panel). Our results are to first order in tension with the recent simulation study by \mbox{\citet{Teklu17}}, who find good agreement with the measured DMFs in \mbox{\citet{Genzel17}}, i.e. smaller than 20~per~cent. Their $z=2$ galaxies, both disc-dominated and bulge-dominated, exhibit circular velocity curves that are somewhat flatter than ours. However, they also perform virtual observations to obtain {\it rotation} curves, and these fall more steeply with radius (50~per~cent within 10~kpc) than do our circular velocity curves (20~per~cent). They put the difference between their rotation and circular velocity curves down to kinetic pressure on the tracer gas. It may therefore be the case that some combination of the matter distribution and gas dynamics of the tracers is required to match the observations.

We have checked our results by looking at the whole sample of disc-dominated galaxies and defined whether our curves are `peaked' or `falling' using a similar criterion to that of \citet{Genzel17}, that the peak value of the velocity curve within 10~kpc, $V_\rmn{max}(<10~\rmn{kpc})$, is significantly higher than its value at 10~kpc $V_\rmn{c}(10~\rmn{kpc})$ (3D, physical in our case). A galaxy is therefore considered peaked if $V_\rmn{c}(10~\rmn{kpc})/V_\rmn{max}(<10~\rmn{kpc})<1$. We find that massive galaxies ($V_\rmn{DM+b}>200$\kms) are almost always peaked at $z=2$, compared to 60~per~cent peaked for $z=0$. However, the strength of the peaks, as defined by the ratio of $V_{c}(10~\rmn{kpc})$  to $V_\rmn{max}(<10~\rmn{max})$, is stronger at $z=0$, and particularly for less massive galaxies ($V_\rmn{DM+b}<200$\kms) by about 10~per~cent in the ratio. We therefore predict that future studies of fainter high redshift galaxies will yield flatter circular velocity curves than their high mass counterparts. 

We  hence conclude that the TNG simulations predict that high-redshift, massive galaxies have occasionally falling rotation curves and lower DMFs within their effective radii than their low-redshift counterparts, in broad agreement with observations. However, the precise magnitude of the DMFs is still under heated debate: the TNG simulations return larger DMFs than those reported by many but not all observational studies, this possibly being due to an excessive contraction of the underlying dark matter haloes. However, more work is required to solve the apparently anomalous and inconsistent results for DMFs inferred for observed galaxies at low-redshifts, including our own Galaxy.

   \section{Summary and Conclusions}
	\label{con}
    
The presence of dark matter in galaxies has become part of the paradigm of cosmology and galaxy formation, providing a valuable theoretical framework to understand the phase-space properties of galaxies of all masses, sizes and morphological types. However, the precise nature of the relationship between dark matter and galaxies, and hence the exact amount of dark matter within the bright bodies of galaxies, how it interacts with stars and gas, and the resulting radial distributions of all matter components within galaxies remain open questions. Recent developments in microlensing observations \citep{Wegg16}, kinematic tracers  \citep{Cappellari13,Wojtak13,Zhu14,Alabi17,ZhuL18}, dynamical models (\mbox{\citealp{Thomas07,Tortora09}},\citealp{Humphrey10}; \mbox{\citealp{Toft12,Tortora12,Beifiori14}};  \mbox{\citealp{Tortora14,Tortora16a,Tortora16b,Oldham18}}) and gravitational lensing \citep{Koopmans06,Tortora10,Barnabe11,JimenezVicente15} are probing the distribution of matter in galaxies with increasing precision, but large systematic discrepancies and possibly intrinsic variations still remain across the mass fractions of  dark matter inferred from observations.

In this paper we have furthered our understanding of this picture from a theoretical perspective and analyzed the galaxy population of TNG100 and TNG300, the first two completed simulations of the IllustrisTNG (hereafter TNG) project. This is a series of calculations that include a comprehensive set of baryonic processes that are relevant for the formation and evolution of galaxies and their gaseous and dark matter haloes. The TNG100 and TNG300 cosmological volumes together allow us to model four orders of magnitude in stellar mass and seven orders of magnitude in halo mass. Their galaxy populations have been shown to be in reasonable agreement with a host of observational constraints -- of particular relevance for our analysis, the abundance and mass of galaxies and their stellar sizes at low redshifts. 

We therefore have an excellent dataset with which to make predictions for the matter distribution and dark matter content of galaxies across multiple galaxy masses and types. In particular, in this paper we have analysed the dark matter mass fraction (DMF) within different galaxy apertures and the circular velocity curves of objects at both $z=0$ and around the peak of star formation activity, $z\sim2$, by comparing the TNG haloes to their dark-matter only (DMO) counterparts, with the goal of providing a theoretical benchmark to recent and upcoming observational analysis.\\

From the inspection of the galaxy population produced with the TNG model and many of its variations, we first of all notice that the effective DMFs within galaxies not only depend on how the baryonic mass is spatially distributed, but also on how baryonic physics may alter the distribution of dark matter itself in comparison to dark matter-only expectations. Therefore, for any given numerical model, the resulting DMFs within a certain galaxy aperture, and their trend with galaxy stellar mass, depend on the {\it complex} combination of: a) the shape and magnitude of the galaxy stellar mass to halo mass (SMHM) relation; b) the shape and magnitude of the galaxy size to galaxy/halo mass relation; and c) the degree to which the dark matter distribution is modified by baryonic processes in comparison to what is expected for NFW haloes. Specifically, at fixed halo mass, larger galaxy masses and smaller galaxy sizes imply lower DMFs; dark matter halo contraction enhances the DMF values. 
More quantitatively, our results can be summarized as follows:  

\begin{itemize}

\item The diverse baryonic mass distributions within galaxies translate into diverse dark matter fractions measured within multiples of the stellar sizes, across galaxy masses, types and also at fixed mass. Within the 3D stellar half-mass radius at $z=0$, the DMF is not a monotonic function of galaxy mass. The TNG model predicts median dark matter mass fractions, $f_{\rm DM}(< r_{\rm half})$, larger than 50 per cent at all galaxy masses, with a minimum for MW-mass galaxies: $f_{\rm DM}(< r_{\rm half}) = 0.5 \pm 0.1$. This budget increases to $0.7-0.8$ within five times the stellar half mass radius, and to more than 80 per cent  dark matter domination for dwarf galaxies ($M_{*}<10^{9}\Msun$, Fig.~\ref{fDM}).\\

\item Most of the scatter in the DMF-stellar mass relation results from the scatter in the sizes of galaxies when computing DMFs within $r_\rmn{half}$, and to a lesser degree from the DMFs within an aperture of fixed size. Namely, more compact galaxies (the lowest quartile in $r_\rmn{half}$) at fixed stellar mass have up to 20 per cent lower DMFs than average-size galaxies, and vice-versa for more extended galaxies (Fig.~\ref{fDM2}, left). There is a correlation between size and morphological type -- elliptical galaxies are on average smaller than disc-like galaxies of the same stellar mass -- and therefore elliptical galaxies also have lower DMFs (Fig.~\ref{fDM2}, right). We find that this scatter is also influenced by the properties of the host halo, even at fixed stellar mass, and that the scatter in the least-massive galaxies is strongly related to their star formation rate. \\

\item The evolution with time of the dark matter fractions (Fig.~\ref{RHvZPlus}) {\it de facto} depends on the aperture within which they are measured. In TNG, the DMF within $r_\rmn{half}$ increases by up to about 50 per~cent from redshift 2 to redshift 0, at least for massive galaxies ($\ga 10^{11}\Msun$). However, this evolution is mainly due to the increase in $r_\rmn{half}$ itself: the DMF within a fixed aperture of 5~kpc or for other stellar masses instead is essentially flat in both models, and actually decreases slightly towards $z=0$ for $10^{10}\Msun$ galaxies (Fig.~\ref{RHvZPlus}). At $z\sim2$, TNG galaxies for the stellar mass range $10^9 - 10^{11}\Msun$ are predicted to have dark matter fractions ($< r_{\rm half}$) in the range $0.35-0.8$ (Fig.~\ref{RHvZPlus}, bottom panel).\\

\item Interestingly, the TNG simulations would predict up to a factor of two lower DMFs at all times {\it if} they did not also predict an enhancement of the dark matter mass within the galaxy bodies in comparison to DMO expectations (Figs.~\ref{fDM}, \ref{SLUGGS} and \ref{GENZEL}), such an enhancement or dark matter halo contraction being an emergent feature of our model. In other words, TNG dark matter fractions are larger than those that would have been expected by placing the baryonic TNG galaxies inside their DMO halo counterparts, pointing towards an overall contraction of the underlying dark matter haloes (Fig.~\ref{VmaxRmax}). \\

\item Finally, the TNG and the previous Illustris models {\it de facto} predict similar DMFs within the stellar half-mass radius (or a few kpc in galactocentric distance) as a function of galaxy stellar mass, at $z=0$ (Fig.~\ref{fDM} vs. Fig.~\ref{MODELS}). However, this inter-model consistency is somewhat coincidental, as at fixed halo mass Illustris galaxies are more massive and more extended than their TNG counterparts and in worse agreement with observational constraints than TNG galaxies. In fact, at least for MW-like galaxies, the two models make very different predictions in the shape of the circular velocity curve, with TNG flatter velocity profiles in better agreement with our Galaxy observations (Fig.~\ref{BOVY}).

\end{itemize}

To put our results consistently into the context of recent observational measurements of the dark matter content of galaxies is not a straightforward task, as it would require us to carefully select the simulated galaxies to be contrasted to the observational samples and to retrace the steps of the observational analyses on virtual or mock observations of our simulated galaxies. Moreover, even for our own Galaxy, the DMFs determined from observations may span very large value ranges (see Table~\ref{ObsTab}), making general statements inappropriate. However, in this work we have attempted to juxtapose the theoretically-derived DMFs to those inferred from observations at face value,  at both the current time for MW-like galaxies (Fig.~\ref{BOVY}) and elliptical ones (Fig.~\ref{SLUGGS}), as well as disc-like galaxies across times (Fig.~\ref{GENZEL}).

\begin{itemize}

\item  For MW-like galaxies at $z=0$, TNG predicts flat {\it total} circular velocity curves beyond a few kpc from the galaxy centre, in agreement with all observational constraints. However, the decomposition into baryonic vs. dark material is more controversial. The DMFs predicted by the TNG model are significantly higher than those allowed by the observational constraints of \citet{Bovy13}: we have checked that among the many hundreds of MW-like galaxies in the TNG100 box, none of them can simultaneously satisfy the \citet{Bovy13} total and dark matter inferred profiles. The tension with the results of \citet{Wegg16} and \citet{IPB15} (see Fig.~\ref{BOVY} and Table~\ref{ObsTab}) is  significantly less severe (depending on aperture), yet these models still prefer dark matter density slopes that are NFW rather than contracted. Our DMFs fall among the estimates for generally disc-like galaxies at $z=0$ from the two DiskMass and SWELLS samples analysed by \mbox{\citet{Courteau15}} and the \citet{Kranz03,Fragkoudi17} data -- see Fig.~\ref{GENZEL}. In fact, our studies demonstrate that the form of the IMF is vital to whether any agreement is achieved, with Salpeter IMF-based DMFs being up to 40~per~cent smaller than those obtained by assuming or preferring a Chabrier IMF. We also note that the total mass profiles of MW-like galaxies from the original Illustris simulation rise too steeply with radius, and hence the Illustris prediction fails to match the observational constraints for any compatible MW halo mass.\\

\item We have shown that TNG elliptical galaxies exhibit DM fractions measured within $5 \times r_\rmn{half}$ higher than those inferred for local elliptical galaxy measurements made in the {\sc sluggs} survey \mbox{\citep{Alabi17}} but are overall in the ball park of those by \citet{Wojtak13}. The intrinsic galaxy-to-galaxy variation in TNG results (10 per cent, 1$-\sigma$) is much smaller than the scatter in the observations, that reaches about 40 per cent, accounting for both an intrinsic and a measurement component (Fig.~\ref{SLUGGS}, top left). A clear tension is apparent when comparing TNG results with the very low DMFs measured within the central regions of the {\sc atlas$^\rmn{3D}$} \citep{Cappellari13} elliptical galaxies, at $1\times r_\rmn{half}$ (Fig.~\ref{SLUGGS}, bottom left). Yet, other data analysed with other techniques (and additionally that assume less bottom-heavy IMFs i.e. an IMF consistent with the simulations: Chabrier) exhibit reasonably good agreement \citep[e.g.][Fig.~\ref{SLUGGS}, bottom right]{Barnabe11,Tortora12}. The SLUGGS vs. TNG inconsistencies are resolved when dark-matter fractions are measured consistently within the same aperture values, e.g. when adopting fixed apertures (30, 40, 50 kpc) rather than multiples of $r_\rmn{half}$, pointing to a residual tension in the inferred galaxy sizes. Other possible sources for discrepancy may include excessive contraction of the halo, the necessity of performing more careful virtual observations, the modelling of steeper dark matter profiles in observations, or a combination of all three.\\

\item At high redshifts, TNG disc-like galaxies  have average DMFs larger by a factor of $\la2$ than those inferred by \mbox{\citet{Genzel17}}: in our model, this seems mostly due to the underlying contraction of dark matter haloes in comparison to DMO predictions, while agreement may be excellent if TNG galaxies were sitting at the centre of {\it DMO} haloes, i.e. if contraction of the halo did not occur in the TNG framework. However, the TNG simulations predict that high-redshift, massive galaxies have occasionally falling circular velocity curves and lower DMFs within their effective radii than their low-redshift counterparts, in broad agreement with observational findings. 

\end{itemize}

In conclusion, we have used the TNG100 and TNG300 simulations as effective models for galaxy formation to quantify the dark matter content and distribution in both local and high redshift galaxies. The prescriptions for AGN and star formation wind feedback succeed in broadly matching the galaxy stellar mass function at low redshifts, the shape of the (total) circular velocity curve of MW-like galaxies, and the evolution of the DMF with redshift, although apparently not the exact amplitude at each redshift. The TNG simulation model returns larger DMFs than those reported by some but not all observational studies. Possible tensions with observations include insufficient scatter in DMFs at $z=0$ and also an excess of dark matter in the very centres of galaxies, {\it if} simulated and observational results can be compared at face value. Much of the apparent tension between simulations and observations is alleviated or completely reduced once a consistent IMF is adopted in both simulated and observational data and if the dark-matter fractions are measured within the same aperture values. We advocate for future DMFs constraints to be provided within a set of fixed physical apertures in kpc, even if across wide ranges of masses, in order to at least short circuit the additional uncertainties associated to the estimates of the galaxy sizes. We anticipate that future work, combining observational work, galaxy dynamical modelling and virtual observations, will help us to learn whether our models are indeed in good agreement with nature, or that instead the dark matter density profile has the potential to tell us something about how feedback unfolds in galaxies.    
  
 \section*{Acknowledgements}

We would like to thank Duncan Forbes, Andreas Burkert, Ling Zhu, Matthieu Schaller, Philipp Lang, and Aaron Romanowsky for inspiring conversations, and we would also like to thank the anonymous referee for a very careful and detailed report. We are very grateful to the whole SLUGGS team for making the fixed aperture measurements of SLUGGS dark matter fractions available to us and for a great deal of valuable input.  MRL is  supported  by  a  COFUND/Durham  Junior Research Fellowship under EU grant 609412. MRL acknowledges
support by a Grant of Excellence from the Icelandic Research Fund (grant number 173929$-$051). VS, RP, and RW acknowledge support through the European Research Council under ERC-StG grant EXAGAL-308037, and would like to thank the Klaus Tschira Foundation. The Flatiron Institute is supported by the Simons Foundation. The simulations used in this work were run on the Hydra and Draco supercomputers at the Max Planck Computing and Data Facility (MPCDF, formerly known as RZG) in Garching near Munich and on the HazelHen Cray XC40-system at the High Performance Computing Center Stuttgart as part of project GCS-ILLU of the Gauss Centre for Supercomputing (GCS).

  \bibliographystyle{mnras}

\label{lastpage}

\end{document}